\documentclass[11pt,a4paper]{article}

\usepackage{amsmath}
\usepackage{amssymb}
\usepackage{graphicx}
\usepackage{dcolumn}
\usepackage{bm}
\usepackage{tabularx}
\usepackage[titles]{tocloft}
\usepackage{cite}
\usepackage{url}
\usepackage{xspace}
\usepackage{csquotes}
\usepackage{xcolor}

\title{Cosmological scalar fields and Big-Bang nucleosynthesis}
\author{A. Arbey, J.-F. Coupechoux}

\begin{document}

\begin{flushright}
CERN-TH-2019-111
\end{flushright}

\begin{center}
\Large\bf\boldmath
\vspace*{2.0cm}Cosmological scalar fields and Big-Bang nucleosynthesis
\unboldmath
\end{center}

\vspace{0.8cm}
\begin{center}
A. Arbey$^{a,b,}$\footnote{Electronic address: \tt alexandre.arbey@ens-lyon.fr}$^,$\footnote{Also Institut Universitaire de France, 103 boulevard Saint-Michel, 75005 Paris, France}, J.-F. Coupechoux$^{a,}$\footnote{Electronic address: \tt j-f.coupechoux@ipnl.in2p3.fr}\\[1.cm]

{\sl $^a$Univ Lyon, Univ Lyon 1, CNRS/IN2P3, Institut de Physique Nucl\'eaire de Lyon UMR5822, F-69622 Villeurbanne, France.\\[0.5cm]}

{\sl $^b$Theoretical Physics Department, CERN, CH-1211 Geneva 23, Switzerland\\[0.2cm]}

\end{center}
\vspace{1.5cm}
\begin{abstract}
Scalar fields are widely used in cosmology, in particular to emulate dark energy, for example in quintessence models, or to explain dark matter, in particular within the fuzzy dark matter model. In addition many scenarios involving primordial scalar fields which could have driven inflation or baryogenesis are currently under scrutiny. In this article, we study the impact of such scalar fields on Big-Bang nucleosynthesis and derive constraints on their parameters using the observed abundance of the elements.
\end{abstract}

\newpage

%%%%%%%%%%%%%%%%%%%%%%%%%%%%%%%%%%%

\section{Introduction}

The nature of dark energy and dark matter is still an unresolved mystery of cosmology. The questions of baryon asymmetry in the Universe, baryogenesis and inflation also necessitate new phenomena in the early Universe. Many cosmological models for such phenomena involve scalar fields. In addition, with the discovery of a Standard Model Higgs boson \cite{Aad:2012tfa,Chatrchyan:2012xdj} the existence of fundamental scalar fields has been proven.

The current cosmological observations can be explained by assuming the existence of cold dark matter (CDM) and of a cosmological constant $\Lambda$ \cite{Aghanim:2018eyx}, forming the cosmological standard model $\Lambda$CDM. The cosmological constant currently represents about two-third of the present total energy density in the Universe, whereas CDM and baryons constitute the remaining energy density. CDM is a pressureless component, and the cosmological constant has a negative pressure and constant density. At the time of recombination and emission of the cosmic microwave background (CMB), the dominating energy density was that of radiation, and the baryon and CDM densities were subdominant but left specific imprints on the CMB, and the cosmological constant was completely negligible. Before recombination, radiation is assumed to dominate, and in the very early Universe the expansion is expected to be exponential during the inflation period.

Typical inflation models \cite{Linde:1981mu,Albrecht:1982mp} involve a scalar field, the inflaton, which is dominated by its potential, leading to a constant scalar field density, which decayed at a later stage into Standard Model particles.

Similarly, the cosmological constant can be replaced by a dark energy with a nearly constant density today \cite{Zlatev:1998tr,Amendola:1999er,Caldwell:1999ew,Chiba:1999ka,Bento:2002ps,Nojiri:2005pu,Tsujikawa:2013fta}. Contrary to the cosmological constant, dark energy is driven by a dynamical mechanism. Quintessence models for example aim at explaining dark energy with a cosmological field \cite{Zlatev:1998tr,Amendola:1999er}. The form of its potential is clearly unknown, and many different potentials have been studied and confronted to observations, and a large variety of models are still compatible with observational data.

Typical dark matter models involve weakly-interacting massive particles (WIMPs), which can be scalar particles still undiscovered at colliders and dark matter detection experiments. Models for dark matter can also be based on other kinds of scalar fields. This is for example the case of fuzzy dark matter \cite{Hu:2000ke} or spintessence \cite{Boyle:2001du}, in which the scalar field oscillates quickly, acquiring in average a pressureless matter behaviour. At galactic scales, is can form a galactic halo through Bose-Einstein condensation, whose typical size is given by the Compton wavelength of the scalar field. Such models need a quadratic term in the potential with mass as low as $10^{-23}$ eV, and therefore the scalar field does not behave like particles. Such models have the advantage of not having the so-called cuspy halo and missing satellite problems \cite{Irsic:2017yje}.

On the other hand, the dark fluid model \cite{Peebles:2000yy,Bilic:2001cg,Arbey:2005fn} describe a cosmological component which behaves at galactic scales as dark matter and at large scales like dark energy, replacing both components by a single one. Such a model can rely on a scalar field with a specific potential \cite{Arbey:2006it,Arbey:2007vu,Arbey:2008gw}.

Other scalar fields could have appeared in the early Universe, for example being the manifestation of extra-dimensions (moduli \cite{Dine:1995uk,Banks:1995dt,Moroi:2001ct,Nakamura:2006uc}, dilatons \cite{deCarlos:1993wie,Gasperini:1994xg}, ...) or high energy theories. They may have driven phenomena such as inflation or leptogenesis, and are expected to have decayed much before recombination.

In this article, we study the effects of cosmological scalar fields at the time of Big-Bang nucleosynthesis (BBN), where they could have affected the formation of heavy nuclei. We obtain constraints on scalar field models confronting the obtained abundance of the elements to observational constraints. We also compute the abundance of lithium-7 and show that cosmological scalar field cannot help solving the so-called lithium problem \cite{Coc:2003ce}.

We restrict our study to scalar fields with standard kinetic terms and positive potentials within the cosmological standard model, therefore disregarding more exotic cases such as Chaplygin gas \cite{Bilic:2001cg,Bento:2002ps}, phantom energy \cite{Caldwell:1999ew,Nojiri:2005pu,Capozziello:2005tf} or $k$-essence \cite{Piazza:2004df,Tsujikawa:2004dp}. We also do not consider scalar particles, but only scalar fields which have a large Compton wavelength.

The article is organized as follows: we derive first the master equations for the evolution of the scalar field, and present the BBN equations and observational constraints. We then study the evolutions of scalar fields of dark energy, dark matter and dark fluid. We finally derive constraints on such models using BBN constraints.

%%%%%%%%%%%%%%%%%%%%%%%%%%%%%%%%%%%

\section{Master equations}

In this section, we establish the equations which describe the evolution of a scalar field during BBN and which will allow us to compute the abundance of the elements and to set limits on the scalar field scenarios using observational constraints. In the following, we will use natural units with $c=\hbar=1$.

The cosmological parameters needed for our study are summarized in Table~\ref{tableau}.

\begin{table}[!ht]
\begin{center}
 \begin{tabular}{|c|c|c|}
 \hline Parameter & Definition  & Observational value \\
 \hline	$\omega_b=\Omega^0_bh^2$ & Baryon cosmological parameter  & $0.02237\pm 0.00015$\\
 \hline $\omega_c=\Omega^0_ch^2$ & Cold dark matter cosmological parameter  & $0.1200\pm 0.0012$\\
 \hline $\Omega_{\Lambda}$ & Cosmological constant parameter  & $0.6847 \pm 0.0073$\\
 \hline  $H_0$ & Hubble constant in km/s/Mpc & $67.36\pm 0.54$\\
 \hline  $z_{eq}$ & Redshift of matter-radiation equality & $3391\pm 60$\\
 \hline  
 \end{tabular}
 \caption{Main cosmological parameters \cite{Aghanim:2018eyx} used in our study. $\Omega^0_X$ is defined as the ratio of the current energy density of component X over the present critical density, and $h$ is the reduced Hubble constant $h=H_0/(100$ km/s/Mpc).\label{tableau}}
\end{center}
\end{table}

\subsection{Cosmological scalar field}

We consider a real and neutral scalar field $\phi$ described by the following Lagrangian density
\begin{equation}
\mathcal{L}_{\phi}=\frac{1}{2}g^{\mu\nu}\partial_{\mu}\phi\partial_{\nu}\phi+U(\phi) \,,
\label{L}
\end{equation}
where the first term is the usual kinetic term, $g^{\mu\nu}$ the inverse metric and $U$ the potential of the scalar field defined positive. The total action is given by \cite{Seidel:1990jh}:
\begin{equation}
\mathcal{S}=\frac{1}{16\pi G}\int d^4x\sqrt{-g}R-\int d^4x\left[\sqrt{-g}\mathcal{L}_E\right]\,,
\label{action}
\end{equation}
where $G$ is the Newton constant, $g$ the determinant of the metric, $R$ the scalar curvature and $\mathcal{L}_E$ the total energy Lagrangian density.

The variation of the action with respect to the scalar field gives the Klein-Gordon equation:
\begin{equation}
\Box \phi = \frac{1}{\sqrt{-g}}\partial_{\mu}\sqrt{-g}g^{\mu \nu}\partial_{\nu}\phi= \frac{dU}{d\phi}\,,
\label{eq KG}
\end{equation}
and the variation with respect to the metric the Einstein equations:
\begin{equation}
G_{\mu\nu}=8\pi GT_{\mu\nu} \,,
\label{eq E}
\end{equation}
where $G_{\mu\nu}$ is the Einstein tensor and $T_{\mu\nu}$ is the total stress-energy tensor. The contribution of the scalar field to the stress-energy tensor is given by:
\begin{equation}
T_{\mu\nu}^\phi=\frac{2}{\sqrt{-g}}\frac{\delta \left(\sqrt{-g}\mathcal{L}_{\phi}\right)}{\delta g^{\mu\nu}}=\partial_{\mu}\phi\partial_{\nu}\phi - g_{\mu\nu}\mathcal{L}_{\phi}\,.
\end{equation}

The previous equations are valid in any metric. To study the dynamics of a scalar field in a homogeneous background Universe, we adopt the Friedmann-Lemaître-Robertson-Walker (FLRW) metric, assuming a flat geometry, such that $d\tau^2=-dt^2+a^2(t)d\vec{r}^{\,2}$ where $a(t)$ is the scale factor at cosmological time $t$ and $\vec{r}$ are the comoving coordinates. Within this metric, the Klein-Gordon and Einstein equations become:   
\begin{equation}
\begin{aligned}
&H^2=\frac{8\pi G}{3}(\rho_{\phi} + \rho_{\rm other})\,,\\
&2\dot{H}+3H^2=-8\pi G \, (P_{\phi}+ P_{\rm other})\,,\\
&\Ddot{\phi}+3H\dot{\phi}+\frac{dU}{d\phi}=0\,,\label{KG_E}
\end{aligned}
\end{equation}
where the energy density and pressure of the scalar field are given by
\begin{equation}
\begin{aligned}
&\rho_{\phi}=\frac{1}{2}\left(\frac{d\phi}{dt}\right)^2+U(\phi)\,,\\
&P_{\phi}=\frac{1}{2}\left(\frac{d\phi}{dt}\right)^2-U(\phi)\,.\label{densities}
\end{aligned}
\end{equation}
$\rho_{\rm other}$ and $P_{\rm other}$ are the sums of the densities and pressures of the other cosmological components, $H=\dot{a}/a$ is the Hubble parameter, and $a$ is the scale factor. The constant $a_0$ is the present value of the scale factor, that we set in the following to $a_0=1$ to simplify the equations.

The Universe being in adiabatic expansion, the equation of state of the cosmological component $i$ is defined by the parameter
\begin{equation}
 w_i = \frac{P_i}{\rho_i}\,.
\end{equation}

In the standard cosmological model, the different components of the total energy density are radiation, matter and cosmological constant. All of these constituents can be described by perfect fluids with equations of state $w_\gamma=1/3$ for radiation, $w_m=0$ for matter and $w_\Lambda=-1$ for a cosmological constant.

For the scalar field, two specific behaviours can be forecast:
\begin{itemize}
\item $w_\phi = 1$ when the scalar field has a dominant kinetic term $\dot{\phi} \gg U(\phi)$,
\item $w_\phi = -1$ when the scalar field has a dominant potential term $U(\phi) \gg \dot{\phi}$.
\end{itemize}

%%%%%%%%%%%%%%%%%%%%%%%%%%%%%%%%%%%
%%%%%%%%%%%%%%%%%%%%%%%%%%%%%%%%%%%

\subsection{Big Bang Nucleosynthesis}

In the cosmological standard model, before the beginning of BBN the expansion of the Universe is dominated by radiation. The dominating species are photons $\gamma$, electrons and positrons $e^\mp$, baryons $b$, neutrinos $\nu$ and antineutrinos $\bar\nu$, and dark matter $\chi$. In presence of a scalar field, the total energy density and pressure of the primordial plasma can be written as
\begin{equation}
\begin{aligned}
&\rho_{\rm tot} = \rho_\gamma + \rho_{\nu,\bar\nu} + \rho_{\rm b} + \rho_{e^\mp} + \rho_\chi + \rho_\phi\,,\\
&P_{\rm tot} = P_\gamma + P_{\nu,\bar\nu} + P_{e^\mp} + P_\phi\,, \label{densitiesBBN}
\end{aligned}
\end{equation}
where the baryon and dark matter densities can be considered as pressureless.

The link between temperature and time is given by the conservation of the total radiation entropy, namely:
\begin{equation}
\frac{ds_{\rm rad}}{dt} = -3 H s_{\rm rad} \,.
\end{equation}

The different chemical elements are in interaction through nuclear reactions of the type
\begin{equation}
	N_i\,\, ^{A_i}Z_i + N_j\,\, ^{A_j}Z_j + N_k\,\, ^{A_k}Z_k \leftrightarrow N_l\,\, ^{A_l}Z_l + N_m\,\, ^{A_m}Z_m + N_n\,\, ^{A_n} Z_n\,, \label{reactions}
\end{equation}
where the $N_i$ are the number of nuclei $Z_i$ which enter the reaction and $A_i$ their atomic numbers. The evolution of the number of each of the elements is driven by the Boltzmann equations:
\begin{equation}
	\frac{{d}Y_i}{{d}t} = N_i \sum_{j,k,l,m,n} \left( -\frac{Y_i^{N_i} Y_j^{N_j}Y_k^{N_k}}{N_i!N_j!N_k!}\Gamma_{ijk\rightarrow lmn} + \frac{Y_l^{N_l}Y_m^{N_m}Y_n^{N_n}}{N_l!N_m!N_n!}\Gamma_{lmn\rightarrow ijk} \right)\,, \label{eq_Y}
\end{equation}
where $\Gamma_{ijk\rightarrow lmn}$ and $\Gamma_{lmn\rightarrow ijk}$ are the forward and reverse reaction rates of Eq.~(\ref{reactions}).

BBN occurs for scale factor of about $a_{\rm BBN} \sim 10^{-10}$ and a temperature scale of $T_{\rm BBN} \sim 1$ MeV, and the abundance of the elements after BBN can be obtained by integrating simultaneously Eqs.~(\ref{KG_E}), (\ref{densities}) and (\ref{densitiesBBN})--(\ref{eq_Y}). 

%%%%%%%%%%%%%%%%%%%%%%%%%%%%%%%%%%%

\subsection{Observational constraints}

We compute the abundance of the elements using the public code AlterBBN \cite{Arbey:2011nf,Arbey:2018zfh}, which has been modified to incorporate different kinds of scalar fields. We will compare the abundances to the following set of observational measurements \cite{Tanabashi:2018oca}:
\begin{equation}
\begin{aligned}
&Y_p=0.245 \pm \times 0.003\,,\\
&^2H/H=\left(2.569\pm 0.027\right)\times 10^{-5}\,,\\
&^3He/H=\left(1.1\pm 0.2\right)\times 10^{-5}\,.\label{BBNconstraints}
\end{aligned}
\end{equation}
The exclusion is obtained via a $\chi^2$ analysis at 95\% C.L.

In addition, the observations of the lithium-7 abundance give \cite{Tanabashi:2018oca}:
\begin{equation}
^7Li/H=\left(1.6\pm 0.3\right)\times 10^{-10}\,.\label{Li7BBNconstraint}
\end{equation}
Whereas the observations of the previous abundances are compatible with the predictions in the cosmological standard model, the $^7Li$ abundance shows a discrepancy of more than 3$\sigma$ between the observations and the predictions, which are about a factor three too large. We will check in the following analysis whether the presence of a scalar field can explain this discrepancy.

%%%%%%%%%%%%%%%%%%%%%%%%%%%%%%%%%%%
%%%%%%%%%%%%%%%%%%%%%%%%%%%%%%%%%%%

\section{Dark energy scalar fields}
\label{sect:dark_energy}

Cosmological scalar fields are often used to mimic dark energy, and typical models include quintessence \cite{Caldwell:2005tm,Tsujikawa:2013fta}. In this section, we consider such a scalar field, and the cosmological constant is set to zero in the equations and replaced by the scalar field. Since all cosmological observations are currently compatible with a simple cosmological constant, the main features of such scenarios are a scalar field density close to the dark energy value in the present Universe, an equation of state $w_\phi \approx -1$, and a negligible density at the recombination time. We will consider several cases of quintessence scalar fields \cite{Tsujikawa:2013fta}: scaling freezing model, tracking freezing model and thawing model. For each of them, we choose the parameters and initial conditions so that the scalar field density equals the cosmological constant density today.

To solve the evolution equations, we follow \cite{Magana:2012xe} and define the reduced variables: 
\begin{equation}
\begin{aligned}
    &x=\frac{\dot{\phi}}{\sqrt{6}M_PH}\,,\\
    &y=\frac{\sqrt{U(\phi)}}{\sqrt{3}M_PH}\,,
\end{aligned}\hspace*{2cm}
\begin{aligned}
    &u_r=\frac{\sqrt{\rho_r}}{\sqrt{3}M_PH}\,,\\
    &u_m=\frac{\sqrt{\rho_m}}{\sqrt{3}M_PH}\,,\\
    &u_{\phi}=\frac{\sqrt{\rho_{\phi}}}{\sqrt{3}M_PH}\,,
\end{aligned}
\end{equation}
where $\rho_r$, $\rho_m$ and $\rho_\phi$ refer to radiation, matter and dark energy densities, respectively, and $M_P$ is the Planck mass. With such definitions, the background evolution is given by:
\begin{equation}
\begin{aligned}
    &x'=\frac{3}{2}x\left(\Pi-2x\right)+\sqrt{\frac{3}{2}}\lambda y^2\,,\\
    &y'=\frac{3}{2}\Pi y -\sqrt{\frac{3}{2}}\lambda y x\,,
\end{aligned}\hspace*{2cm}
\begin{aligned}
    &u_m'=\frac{3}{2}\left(\Pi-1\right)u_m\,,\\
    &u_r'=\frac{3}{2}\left(\Pi-\frac{4}{3}\right)u_r\,,\\
    &u_{\phi}'=\frac{3}{2}\Pi u_{\phi}\,,\\
\end{aligned}
\end{equation}  
with
\begin{equation}
\begin{aligned}     
    &\frac{3}{2}\Pi=\frac{3}{2}\left(2x^2+u_m^2+\frac{4}{3}u_r^2\right)=-\frac{\dot{H}}{H^2}\,,\\
    &x^2+y^2+u_r^2+u_m^2+u_{\phi}^2=1\,,\label{eq_evolution}
\end{aligned}
\end{equation}  
where the prime denotes a derivative with respect to the logarithm of the scale factor $N=\ln a$ and $\lambda=-M_P U'/U$. This system of equations is particularly useful to obtain fixed point solutions in the evolution of the scalar field.

%%%%%%%%%%%%%%%%%%%%%%%%%%%%%%%%%%%

\subsection{Scaling freezing models}

\begin{figure}[t!]
\begin{center}
\includegraphics[width=11cm]{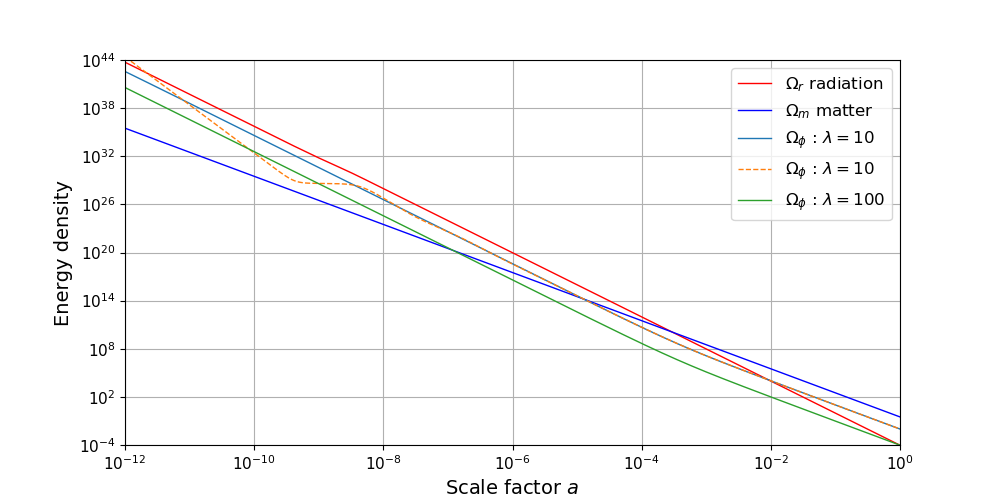}
\caption{Evolution of the scalar field density for an single exponential potential, as a function of the scale factor. The solid curves for the scalar field have been obtained for fixed point initial conditions and the dashed curve for arbitrary initial conditions. In the fixed point case, the evolution of the scalar field density follows the dominant energy density.\label{fig_exp}}
\end{center}
\end{figure}

We first consider the case of a single exponential potential \cite{Copeland:1997et}:
\begin{equation}
    U(\phi)=V\text{exp}\left(-\lambda \frac{\phi}{M_P}\right)\,,
\end{equation}
where $V$ and $\lambda$ are constant parameters. The density evolution of the scalar field is given by the system of equations~(\ref{eq_evolution}). The fixed point solutions for the scalar field respect $x'=0$ and $y'=0$. In the case where the evolution of the Universe is dominated by a barotropic fluid with a pressure such as $P_f=(f-1)\rho_f$ where $f$ is constant, we have: 
\begin{equation}
\begin{aligned}
&x=\frac{\dot{\phi}}{\sqrt{6}M_PH}=\sqrt{\frac{3}{2}}\frac{f}{\lambda}\,,\\
&y=\frac{\sqrt{U(\phi)}}{\sqrt{3}M_PH}=\left(\frac{3(2-f)f}{2\lambda^2}\right)^{1/2}\,,\\
\end{aligned}
\end{equation}
and the fixed point solutions are given by:
\begin{equation}
\begin{aligned}
\frac{8\pi G}{3H^2}\left(\frac{\dot{\phi}^2}{2}+V\text{exp}\left(-\lambda \frac{\phi}{M_P}\right) \right)=\frac{3f}{\lambda^2}\,.
\end{aligned}
\end{equation}
These fixed point solutions are the only ones which are stable, as discussed in \cite{Copeland:1997et}. Therefore during the radiation-domination era we have: 
\begin{equation}
	\left(\frac{H}{H_0}\right)^2 \simeq \Omega_r^0
    \hspace{0.5cm}	
	\Longrightarrow
	\hspace{0.5cm}
	\Omega_{\phi}=\frac{\rho_{\phi}}{\rho^{c}}=\frac{4\Omega^0_r}{\lambda^2} a^{-4}\,,
\end{equation}
and during the matter-domination era:
\begin{equation}
	\frac{H^2}{H_0^2}\simeq \Omega_m^0 a^{-3}
    \hspace{0.5cm}	
	\Longrightarrow
	\hspace{0.5cm}
	\Omega_{\phi}=\frac{\rho_{\phi}}{\rho^{c}}=\frac{3\Omega^0_m}{\lambda^2} a^{-3}\,,
\end{equation}
where $\rho^{c}$ is the critical density. Thus the evolution of the scalar field density will follow that of the dominating density. The evolution of the scalar field is shown as a function of the scale factor in Fig.~\ref{fig_exp}. As expected the scalar field density follows the evolution of the dominating density and is inversely proportional to $\lambda^2$ for fixed point solutions. For $\lambda=10$, two curves are drawn, one with fixed point initial conditions and the other one for arbitrary initial conditions. In the latter case the scalar field is first dominated by its kinetic term and its density is proportional to $a^{-6}$, it then reaches a plateau whose height depends on the value of $V$ and the initial values of $\phi$ and of $\dot{\phi}$. This kind of behaviour does not depend on the form of the potential and will be studied in more detail in the next section. Ultimately, the scalar field reaches the fixed point behaviour.  

At the time of BBN ($a \sim 10^{-10}$), since the radiation is dominating, the scalar field density decreases with an exponent equal to $n_{\phi}=4$ for fixed point solutions and $n_{\phi}=6$ for other initial conditions.

The problem of the exponential potential, however is that, because of its tracking behaviour, it does not lead to a correct present-time behaviour with $w_\phi \approx -1$, and the single exponential potential is generally considered as excluded \cite{Copeland:2006wr}.

\begin{figure}[t!]
\hspace{-2.7cm}
\begin{minipage}[c]{0.63\textwidth}
\includegraphics[width=11cm]{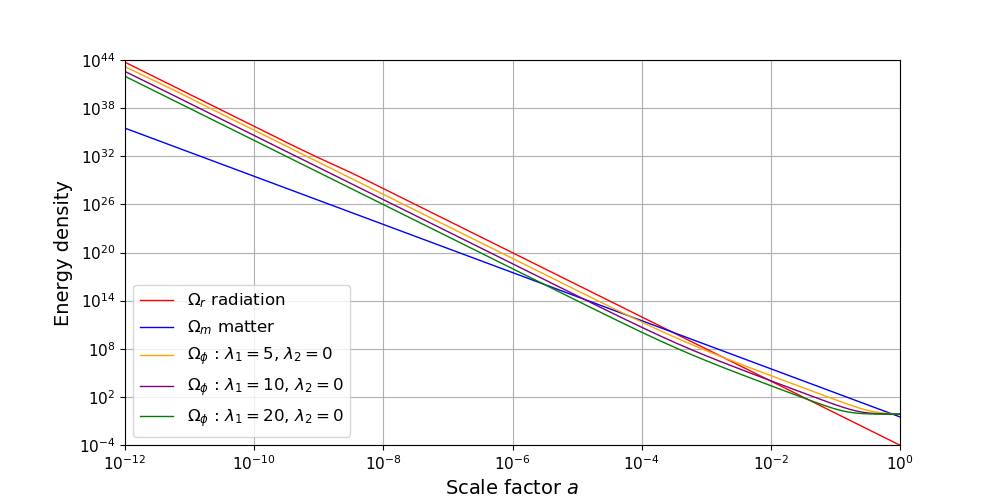}
\end{minipage}
\hspace{0.2cm}
\begin{minipage}[c]{-0.3\textwidth}
\includegraphics[width=11cm]{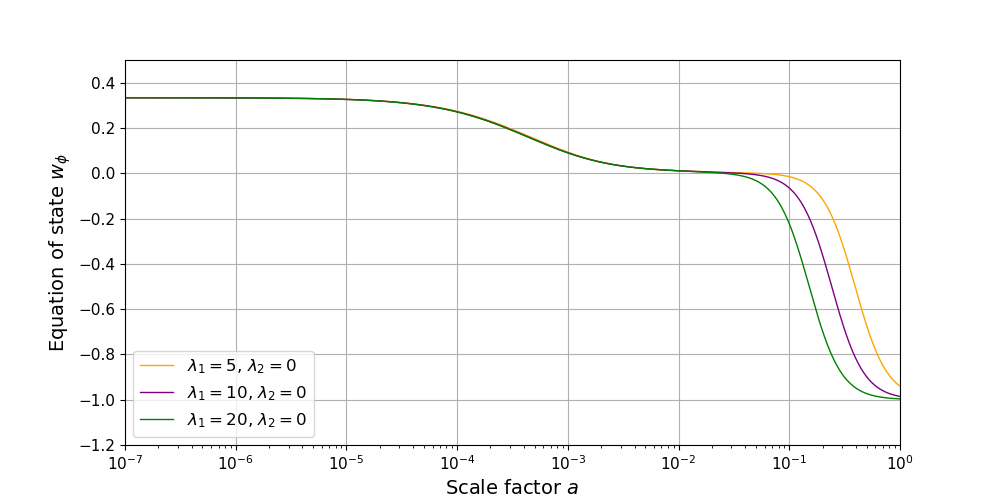}
\end{minipage}
\caption{Evolution of the scalar field in the scaling freezing model with a double exponential behaving like dark energy. $V_1$ and $V_2$ have been fixed to retrieve the cosmological constant value today. The left panel shows the evolution of its density, matter density and radiation density as functions of the scale factor. The right panel shows the evolution of the scalar field equation of state $w_\phi=P_{\phi}/\rho_{\phi}$ as a function of the  scale factor. The curves have been drawn for different values of $\lambda_1$, and for $\lambda_2=0$.\label{fig_exp2}}
\end{figure}

We therefore turn towards the more flexible double exponential potential \cite{Barreiro:1999zs}, which appears as a prototypical scaling freezing model:
\begin{equation}
    U(\phi)=V_1\text{exp}\left(-\lambda_1 \frac{\phi}{M_P}\right)+ V_2\text{exp}\left(-\lambda_2 \frac{\phi}{M_P}\right)\,,
    \label{pot_scaling}
\end{equation}
where $V_{1,2}$ and $\lambda_{1,2}$ are constant parameters. Such a model can be motivated by compactifications in superstring models. The left panel of Fig.~\ref{fig_exp2} shows the evolution of the scalar field density for $\lambda_2=0$ and for different values of $\lambda_1$. The initial conditions have been chosen in order to obtain fixed point solutions. The right panel shows the evolution of the equation of state $w_\phi=P_{\phi}/\rho_{\phi}$. During the radiation-dominated era, we have $w_\phi=w_r=1/3$. During the matter-dominated era we have $w_\phi=w_m=0$. It is only recently that the second term in the potential~(\ref{pot_scaling}) is dominating, leading to an equation of state $w_\phi=-1$. Therefore a double exponential can explain the recent acceleration of the expansion of the Universe.

To summarize, within the model under consideration, at the epoch of BBN the scalar field has a radiation-like behaviour with $w_\phi = 1/3$ for fixed point solutions, whereas with generic initial conditions either the kinetic term dominates leading to $w_\phi = 1$, or the potential dominates leading to a plateau with $w_\phi = -1$.

%%%%%%%%%%%%%%%%%%%%%%%%%%%%%%%%%%%

\subsection{Tracking freezing models}

\begin{figure}[t!]
\hspace{-2.7cm}
\begin{minipage}[c]{0.63\textwidth}
\includegraphics[width=11cm]{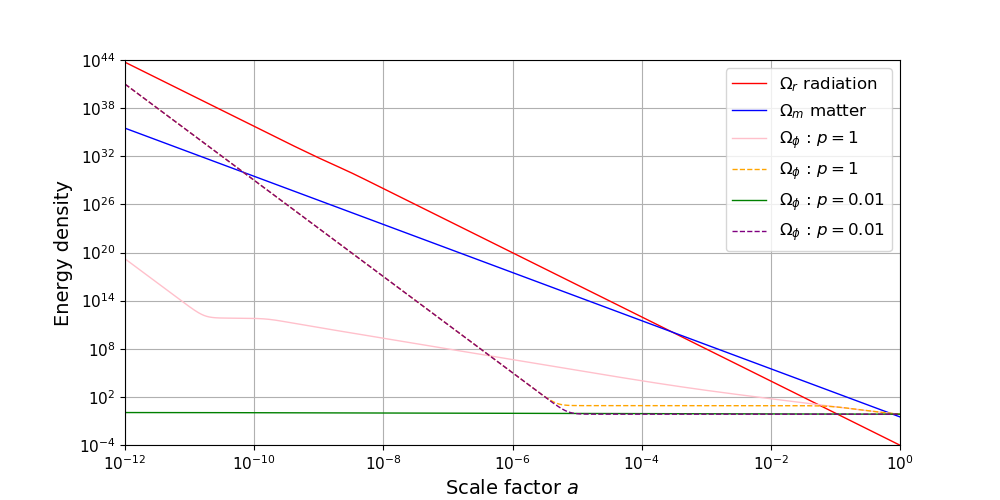}
\end{minipage}
\hspace{0.2cm}
\begin{minipage}[c]{-0.3\textwidth}
\includegraphics[width=11cm]{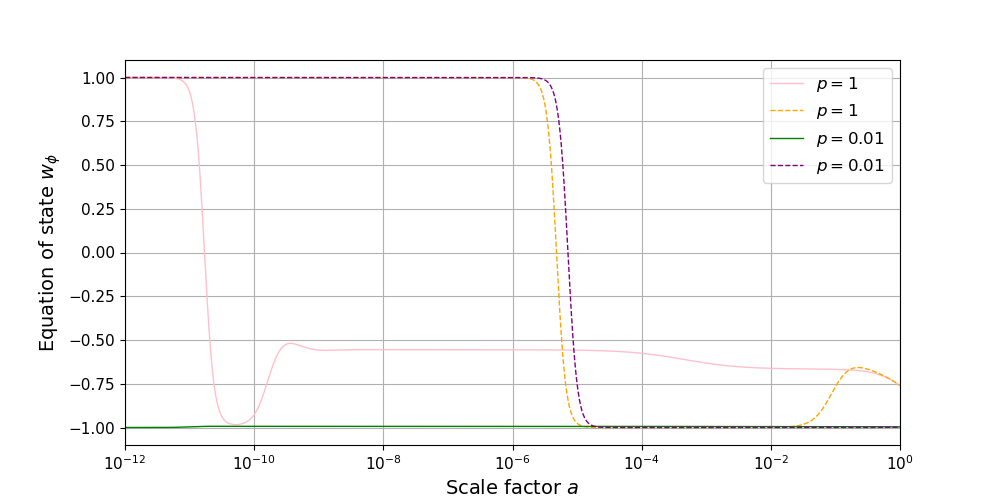}
\end{minipage}
\caption{Evolution of the scalar field in the tracking freezing model with inverse power-low potential behaving like dark energy. $M$ has been fixed to retrieve the cosmological constant value today. The left panel shows the evolution of scalar field, matter and radiation densities as functions of the scale factor. The right panel shows the evolution of the equation of state $w_\phi=P_{\phi}/\rho_{\phi}$ as a function of the scale factor. The curves have been drawn for different values of the exponent $p$ and for two different initial conditions.\label{fig_tracking}}
\end{figure}

We now consider the inverse power-low potential \cite{Linder:2006uf}:
\begin{equation}
    U(\phi)=M^{4}\left(\frac{M_P}{\phi}\right)^p\,,
    \label{pot_power_low}
\end{equation}
where $M$ is a constant mass scale and $p$ a positive exponent. Contrary to the exponential case, the potential is diverging when the scalar field becomes close to zero. 

The left panel of Fig.~\ref{fig_tracking} shows the evolution of the scalar field density for two values of $p$ and different initial conditions. In the early Universe, the scalar field is dominated by its kinetic energy and the equation of state is $w_\phi = 1$. After this period the scalar field density is constant with $w_\phi = -1$. As can be seen in the figure, the duration of the constant behaviour period strongly depends on initial values of scalar field and can even extend to the present period, or stop and reach an intermediate equation of state with $\omega_\phi \sim -0.5$ with a negligible density. Finally the two curves (dashed and solid) for the same value of $p$ are identical and determine the value of the dark energy density today and its equation of state.\\
 
To summarize, in this scenario we found that during BBN the scalar field can either be dominated by its kinetic term with $w_\phi = 1$, or behave like radiation with $w_\phi = 1/3$, or have a negligible density.

%%%%%%%%%%%%%%%%%%%%%%%%%%%%%%%%%%%

\subsection{Thawing models}

\begin{figure}[t!]
\hspace{-2.7cm}
\begin{minipage}[c]{0.63\textwidth}
\includegraphics[width=11cm]{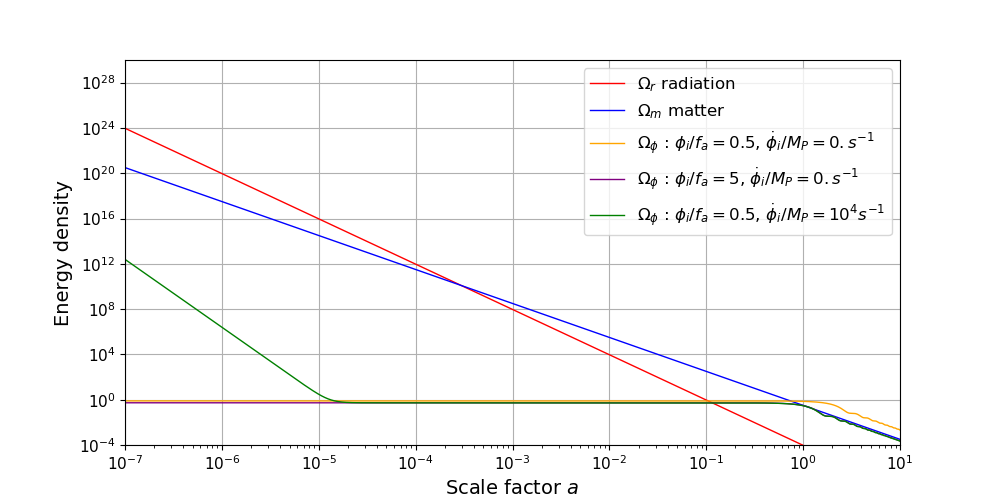}
\end{minipage}
\hspace{0.2cm}
\begin{minipage}[c]{-0.3\textwidth}
\includegraphics[width=11cm]{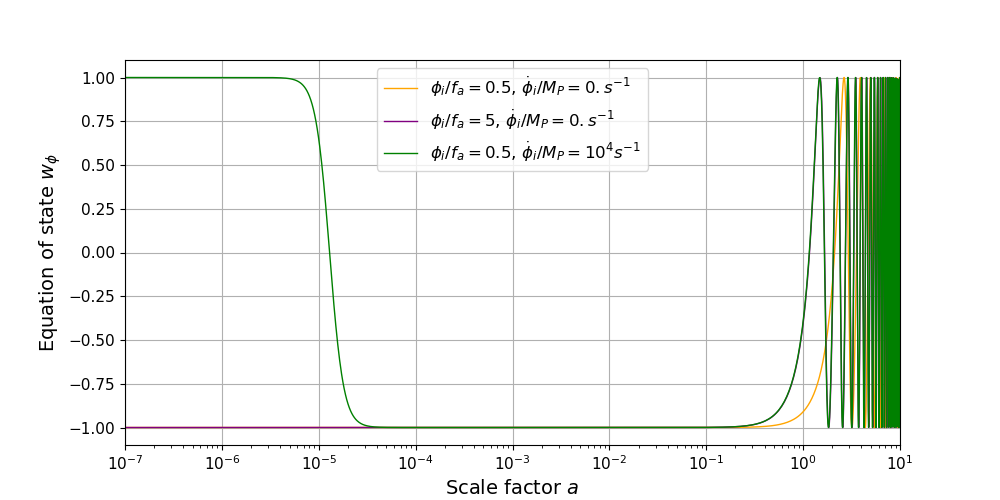}
\end{minipage}
\caption{Evolution of the scalar field in a thawing model with pseudo-Nambu-Goldstone potential behaving like dark energy, for different choices of initial conditions and parameters. $\mu$ has been fixed to retrieve the cosmological constant value today. The left panel shows the evolution of the scalar field, matter and radiation densities as functions of the scale factor. The right panel shows the evolution of the equation of state $w_\phi=P_{\phi}/\rho_{\phi}$ as a function of the scale factor. The curves have been drawn up to $a=10$ for different initial conditions.\label{fig_thawing_init}}
\end{figure}

We now consider the pseudo-Nambu-Goldstone potential \cite{Frieman:1995pm}:
\begin{equation}
    U(\phi)=\mu^4\left(1+\text{cos}(\phi/f_a) \right)\,,
    \label{pot_NG}
\end{equation}
where $\mu$ and $f_a$ are constant parameters.

Figure~\ref{fig_thawing_init} shows the behaviour of the scalar field for different initial conditions and choices of $\mu$ and $f_a$, as a function of the scale factor. We explicitly plotted the evolution up to $a=10$, which corresponds to the future, in order to better visualize the behaviour changes at late time. If the initial value of $\dot{\phi}$ is non-zero, the scalar field is first dominated by its kinetic term. Then the scalar field becomes constant. At the present period, $a\approx 1$, the scalar field starts to evolve and $w_\phi$ increases and starts to oscillate. This behaviour is called ``thawing'' as opposed to the ``freezing'' behaviour of the previous models. For larger values of $\phi_i/M_P$ the field starts oscillating earlier. The averaged value of $w_\phi$ is close to 0, and the scalar field therefore behaves like matter.
At the time of BBN, we observe that the scalar field is generally dominated by its kinetic term leading to $w_\phi = 1$, or has a constant but negligible density.\\

To summarize, quintessence scalar fields at BBN time have generally three different behaviours:
\begin{itemize}
 \item a dominating kinetic term leading to $w_\phi = 1$,
 \item a tracking radiation-like behaviour with $w_\phi = 1/3$,
 \item a constant behaviour with $w_\phi = -1$, which often corresponds to a negligible density.
\end{itemize}
Intermediate behaviours are still possible, but in our set-up they always corresponds to negligible densities during BBN.

%%%%%%%%%%%%%%%%%%%%%%%%%%%%%%%%%%%
%%%%%%%%%%%%%%%%%%%%%%%%%%%%%%%%%%%

\section{Dark matter scalar fields}

In this section, we consider the case of a scalar field with a matter-like behaviour in the present Universe. We therefore have the cosmological constant density set to its observational values, and the cold dark matter density is replaced by the scalar field density. We thus expect the scalar field density to have between recombination and today the same density as the observed cold dark matter one, and a matter-like behaviour corresponding to $w_\phi=0$.

Two separate cases can occur. First, if the scalar field is associated to a mass term with a large mass, dark matter can then be composed of scalar particles. Second, if the mass is very small, the Compton wavelength is large, and the scalar field will have only large scale effects. We consider only the latter case in the following.

Let us consider the case of an oscillating scalar field, similarly to the case of the pseudo-Nambu-Goldstone potential at late times. If the timescale $T$ of the studied phenomena is much longer than the oscillation period of typical frequency $\omega_{e\rm ff}$, but much shorter than the conformal Hubble time $\mathcal{H}^{-1}$, {\it i.e.} $\mathcal{H}^{-1}\gg T\gg \omega_{\rm eff}$, the averaged equation of state is \cite{Cembranos:2015oya}: 
\begin{equation}
    w_\phi=\frac{\langle P_\phi \rangle}{\langle \rho_\phi \rangle}=\frac{\langle \phi'^2/(2a^2)-U(\phi)\rangle}{\langle \phi'^2/(2a^2)+U(\phi)\rangle}=\frac{\langle U'(\phi)\phi -2U(\phi)\rangle}{\langle U'(\phi)\phi +2U(\phi)\rangle}+\mathcal{O}\left(\frac{\mathcal{H}}{\omega_{eff}}\right) \,,
\end{equation}
where the prime corresponds to the derivative with respect to the conformal time $\eta$, defined as $dt=a\,d\eta$, and $\langle...\rangle$ the average over the time interval $T$. For example, if we take a power-law potential $U(\phi)=\lambda |\phi|^n/n$, we can show that $w_\phi \simeq (n-2)/(n+2)$. Then, the conservation of the stress-energy tensor $\rho_{\phi}'+3(1+w_\phi)\mathcal{H}\rho_{\phi}=0$ gives:
\begin{equation}
    \langle \rho_{\phi} \rangle = \rho_0 a^{-3(1+w_\phi)} =\rho_0 a^{-6n/(n+2)}\,.
\label{rho}
\end{equation}
As a consequence, a quadratic term in the potential leads to a matter-like behaviour.

%%%%%%%%%%%%%%%%%%%%%%%%%%%%%%%%%%%

\subsection{Quadratic potential: fuzzy dark matter}

\begin{figure}[t!]
\begin{center}
\includegraphics[width=12cm]{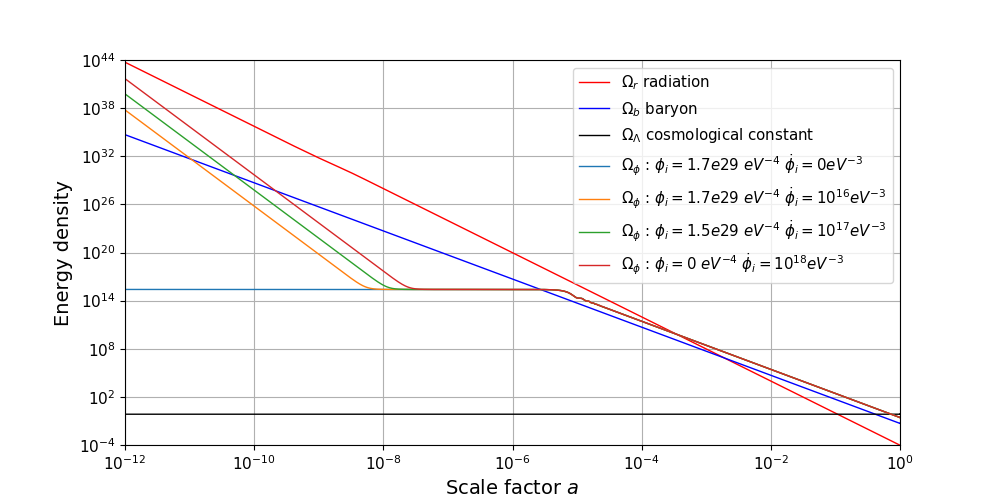}
\caption{Evolution of the fuzzy dark matter scalar field, associated to a quadratic potential with a mass $m=10^{-24}$ eV, as a function of the scale factor. Baryon, radiation and cosmological constant densities follow the standard values of the $\Lambda$CDM model. The density of the scalar field has been calculated for different values of $\dot{\phi}$ at $a=10^{-12}$. For each value of $\dot{\phi}$, we have chosen the initial value of $\phi$ in order to obtain a matter-like behaviour starting before the observed $z_{eq}$ of Table~\ref{tableau} and a present scalar field density equal to the observed cold dark matter density.\label{fig:1}}     
\end{center}
\end{figure}

Fuzzy dark matter \cite{Hu:2000ke} has recently come back into the light \cite{Cembranos:2018ulm} as a potential alternative to WIMP models. It features a quadratic potential such that:
\begin{equation}
    U(\phi)=\frac{1}{2}m^2\phi^2\,.
    \label{U_m}
\end{equation}
As explained before, such a potential can give to the scalar field a matter-like behaviour. Observations of galaxy rotation curves can be used to constrain the value of the mass $m$ \cite{Arbey:2001qi}. At galactic scales, the scalar field can be described as a Bose-Einstein condensate in gravitational interaction with baryonic matter, and a value of $m$ compatible with the observations is of the order of $10^{-24}-10^{-23}$ eV. For the potential (\ref{U_m}) in the homogeneous Universe, the Klein-Gordon equation is given by \cite{Arbey:2001jj}:
\begin{equation}
    \Ddot{\phi}+3H\dot{\phi}+m^2\phi=0\,.
\end{equation}
It is convenient to define the dimensionless time $\Tilde{t}=mt$, the dimensionless Hubble constant $\Tilde{H}=H/m$ and $u=a^{3/2}\phi$. With this change of variables, the Klein-Gordon equation reads:
\begin{equation}
    \Ddot{u}+\left(1-\frac{3}{4}\frac{\dot{a}^2}{a^2}-\frac{3}{2}\frac{\Ddot{a}}{a}\right)u=0\,.
    \label{u}
\end{equation}
Two cases can be considered:
\begin{itemize}
\item assuming $\Tilde{H}\gg 1$, one can transform equation Eq.~(\ref{u}) into 
\begin{equation}
    v''-\frac{a''}{a}v=0\,,
    \label{v}
\end{equation}
with $'=d/d\eta$ derivative with respect to dimensionless conformal time ($d\Tilde{t}=a\,d\eta$) and $v=a^{-1/2}u=a\phi$. In the radiation-domination era, the Friedmann equation gives $a''=0$. Then, we have $d\eta \propto da$ and $v=\alpha_1 a + \beta_1$, where $\alpha_1$ and $\beta_1$ are two constants of integration. The scalar field density reads: 
\begin{equation}
    \rho_{\phi}=\frac{\dot{\phi}^2}{2}+\frac{m^2\phi^2}{2}=m^2\left(\left(\frac{H_0\sqrt{\Omega_r^0}}{a}\frac{d}{da}\left(\frac{v}{a}\right)\right)^2+\frac{1}{2}\left(\frac{v}{a}\right)^2\right)=\frac{m^2H_0^2\Omega_r^0\beta_1^2}{a^6}+\frac{m^2}{2}\left(\alpha_1+\frac{\beta_1}{a}\right)^2\,.
\end{equation}
   
\item assuming $\Tilde{H}\ll 1$ and $\dot{\Tilde{H}}\ll 1$, the solution of Eq.~(\ref{u}) is much simpler. Indeed, the energy $E=\dot{u}^2/2+u^2/2$ is conserved and we obtain
\begin{equation}
    \rho_{\phi}=\frac{\dot{\phi}^2}{2}+\frac{m^2\phi^2}{2}=\frac{m^2E}{a^3}\,.
\end{equation}
Thus the background evolution of scalar field gives a matter-like behaviour.
\end{itemize}

Figure~\ref{fig:1} shows the evolution of the baryon, radiation, cosmological constant and scalar field densities for different initial conditions. In the early Universe the scalar field density is dominated by its kinetic energy and the potential is completely negligible. So $a^3d\phi/dt$ is conserved and the scalar field density is proportional to $a^{-6}$. As $d\phi/dt$ decreases, at some point the field density becomes dominated by the potential and the field density is therefore constant as long as $H>m$. For $H<m$, the scalar field oscillates quickly and its energy density evolves like dark matter.

As discussed earlier, the end of the plateau corresponds to the equality $H^2\simeq m^2$, or equivalently to $a\simeq (H_0^2\sqrt{\Omega_r^0}/m)^{1/4}\simeq 3.7\times 10^{-6}$. Its beginning depends on the initial value of $\dot{\phi}$ and is given by the solution of the equation:
\begin{equation}
    \left(1+\frac{H_0\sqrt{\Omega_r^0}\phi_i}{a_i^2\dot{\phi}_i}\right)a^3-a_i a^2-\frac{a_iH_0\sqrt{\Omega_r^0}}{m}=0\,,
\end{equation}
where the subscript $i$ indicates the initial time. So the plateau begins before $a\simeq 2.4\times 10^{-8}$, which corresponds to a negligible value of $\phi_i$.

In conclusion, because of the plateau, the fuzzy dark matter scalar field density remains negligible during BBN. 

%%%%%%%%%%%%%%%%%%%%%%%%%%%%%%%%%%%

\subsection{Self-interaction coupling}

\begin{figure}[t!]
\hspace{-2.7cm}
\begin{minipage}[c]{0.63\textwidth}
\includegraphics[width=11cm]{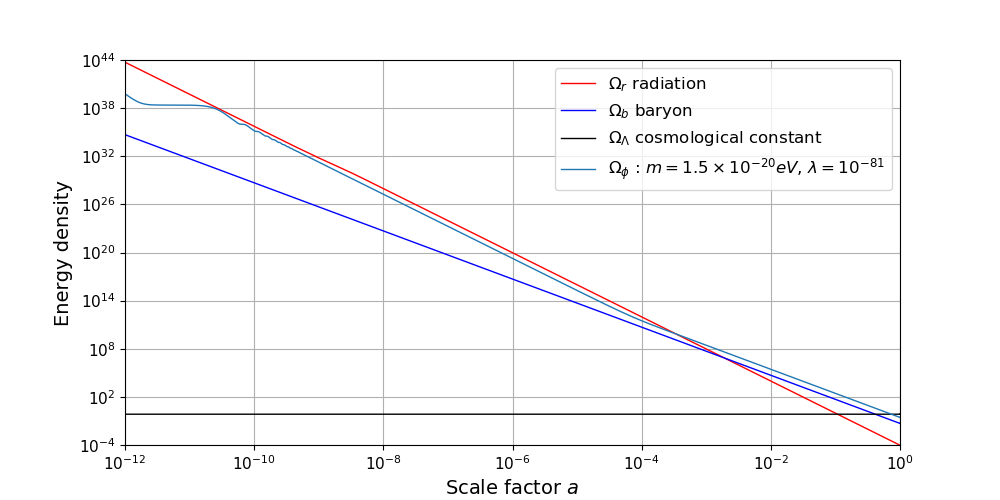}
\end{minipage}
\hspace{0.2cm}
\begin{minipage}[c]{-0.3\textwidth}
\includegraphics[width=11cm]{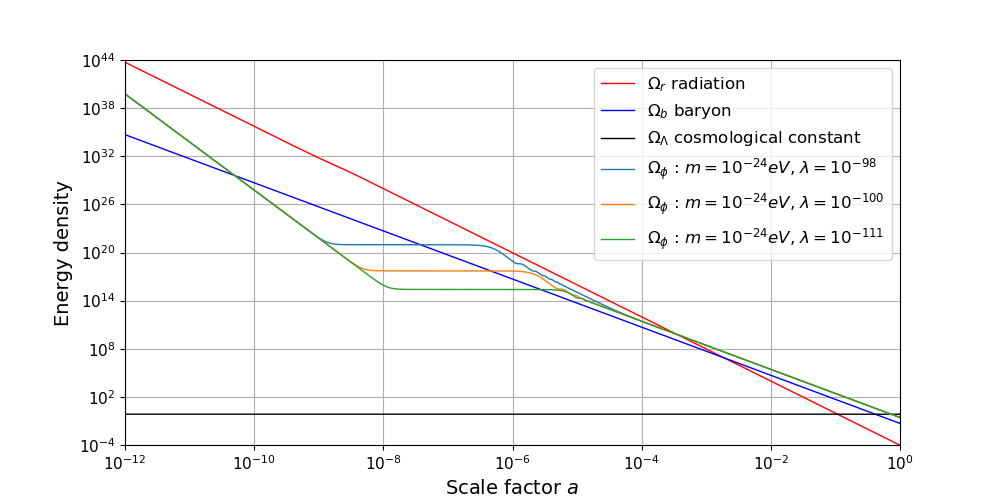}
\end{minipage}
\caption{Evolution of a scalar field with self-interaction coupling behaving like dark matter, as a function of the scale factor. Baryon, radiation and cosmological constant densities follow the standard values of the $\Lambda$CDM model. In the left panel the scalar field density is drawn with $m=1.2\times10^{-20}$ eV, $\lambda=10^{-81}$ and $\phi=9.5\times 10^{29}$ eV$^{-4}$, $\dot{\phi}=1.3\times 10^{17}$ eV$^{-3}$ for $a=10^{-12}$. In the right panel the scalar field density is drawn for different values of $\lambda$. For each value of $\dot{\phi}$, we have chosen the initial value of $\phi$ in order to obtain a matter-like behaviour starting before the observed $z_{eq}$ of Table~\ref{tableau} and a present scalar field density equal to the observed cold dark matter density.\label{fig:2}}
\end{figure}

We extend the previous analysis by adding a quartic term to it \cite{Arbey:2001qi,Arbey:2003sj}:
\begin{equation}
    U(\phi)=\frac{1}{2}m^2\phi^2+\frac{1}{4}\lambda\phi^4\,.
    \label{U_m2}
\end{equation}
The dimensionless constant $\lambda$ represents a self-interaction coupling. With the same notation as in the previous subsection, the Klein-Gordon equation becomes:
\begin{equation}
    v''+\left(a^2 -\frac{a''}{a}+\frac{\lambda}{m^2}v^2\right)v=0\,.
\end{equation}
Three different cases can be considered:
\begin{itemize}
    \item assuming $\lambda\phi^2\ll H^2$ and $\Tilde{H}\gg 1$, the Klein-Gordon equation is the same as Eq.~(\ref{v}) and we obtain similar results.
    
    \item assuming $\lambda\phi^2\gg H^2$ and $\lambda\phi^2\gg m^2$, during the radiation-domination era, the Klein-Gordon equation becomes:
    \begin{equation}
    v''+\frac{\lambda}{m^2}v^3=0\,.
    \end{equation}
    In this case the quadratic part of potential does not contribute to the evolution of field density. The energy $E=v'^2/2+\lambda v^4/(m^2)$ is conserved and we obtain:
    \begin{equation}
    \rho_{\phi}=\frac{\dot{\phi}^2}{2}+\frac{\lambda \phi^4}{4}= \frac{m^2E}{a^4}+\frac{m^2u}{2a^4}\left(\frac{H_0^2\Omega_r^0}{a^2}v-\frac{2H_0\sqrt{\Omega_r^0}}{a}v'\right)\,.
\end{equation}
    Therefore the background behaviour of the scalar field is radiation-like.
    
    \item assuming $\lambda\phi^2\ll m^2$ and $\Tilde{H}\ll 1$, $\dot{\Tilde{H}}\ll 1$, the Klein-Gordon equation is the same as in Eq.~(\ref{u}) and we obtain similar results.
\end{itemize}

Figure~\ref{fig:2} shows the evolution of the baryon, radiation, cosmological constant and scalar field densities. As expected the difference with the results of the previous subsection is a radiation-like behaviour of the scalar field. In the left panel, we have chosen a mass $m=1.2\times 10^{-20}$ eV, and we see that the scalar field density is not negligible during BBN and its evolution is dominated by the self-interaction coupling. Such a mass is however disfavoured by the CMB and large-scale structure data, which impose \cite{Cembranos:2018ulm}:
\begin{equation}
    \begin{aligned}
     & 10^{-26}<m/\text{eV}<10^{-23.3} \,,\\
     & 10^{-111}<\lambda<10^{-98} \,. \label{fdm:constraints}
    \end{aligned}
\end{equation}
In the right panel of Fig.~\ref{fig:2}, we have chosen parameter values compatible with these constraints, and we see that the scalar field density is negligible during BBN. More generally, scanning over the parameters and imposing the constraints~(\ref{fdm:constraints}), we showed that the scalar field density is always negligible at the BBN epoch.

%%%%%%%%%%%%%%%%%%%%%%%%%%%%%%%%%%%

\subsection{Polynomial potential}

The previous results can be extended to the case of an arbitrary polynomial potential. The BBN constraints then depend on which term of the potential is dominating at BBN time. Following Eq.~(\ref{rho}), if the dominating term is $\phi^n$, the scalar field density evolves as $\rho_{\phi} \propto a^{-6n/(n+2)}$, meaning that the exponent is within the range $[-6,0]$. For example, if the dominating term is $\phi^6$, Eq.~(\ref{rho}) gives $\rho_{\phi}\propto a^{-4.5}$. However, since the matter-like behaviour should hold at recombination time, it can be expected that the scalar field density is also negligible at BBN time.

%%%%%%%%%%%%%%%%%%%%%%%%%%%%%%%%%%%

\subsection{Exponential potential}

\begin{figure}[t!]
\begin{center}
\includegraphics[width=12cm]{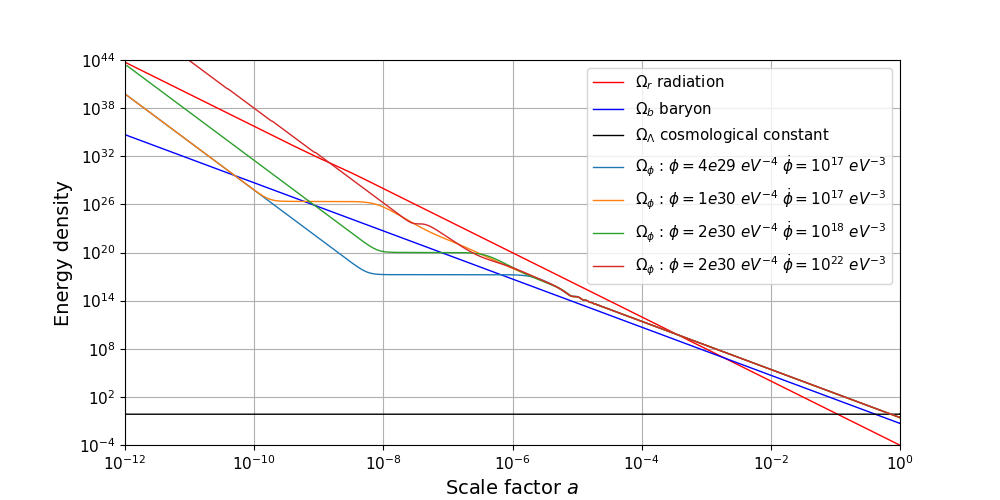}
\caption{Evolution of a scalar field with an exponential potential behaving like dark matter, as a function of the scale factor. Baryon, radiation and cosmological constant densities follow the standard values of the $\Lambda$CDM model. The scalar field density has been calculated for different initial values. For each value of $\dot{\phi}$, we have chosen $\alpha$ in order to obtain a matter-like behaviour starting before the observed $z_{eq}$ of Table~\ref{tableau} and a present scalar field density equal to the observed cold dark matter density.\label{fig::3}}
\end{center}
\end{figure}

We now consider the following potential:
\begin{equation}
    U(\phi)=\alpha \rho^{c}\left[\text{exp}\left(\frac{\beta \phi^2}{m^2}\right)-1\right]\,,
\end{equation}
where $\rho^{c}$ is the critical density, and $\alpha$ and $\beta$ are two dimensionless constants. This potential is built in order to have a minimum at zero in absence of scalar field. Similarly to the previous cases, we want the scalar field to start behaving like dark matter before recombination in order to agree with the CMB data. In the late Universe, the scalar field is expected to have a small value, so that the potential becomes via a Taylor expansion:
\begin{equation}
    U(\phi)=\rho^{c}\left(\frac{\alpha \beta}{m^2} \phi^2+\frac{1}{2}\frac{\alpha \beta^2}{m^4}\phi^4+...\right)\,,
\end{equation}
where the dominant term is quadratic. We therefore recover the fuzzy dark matter potential with possibly a non-negligible quartic term. Constraints on fuzzy dark matter impose the mass of the scalar field to be around $10^{-24}$ eV. In consequence we choose $2\alpha \beta=m^4/\rho^{c}$. With this relation there is only one free parameter for this potential. Figure~\ref{fig::3} shows the scalar field density evolution for different initial conditions. For each initial scalar field value we have chosen $\alpha$ to obtain the same value of the radiation-matter equality as in the $\Lambda$CDM scenario, so that $\alpha\simeq 5\times 10^{15}$. We found that with an exponential potential, only two behaviours are possible during BBN: either a kinetic term domination with $w_\phi = 1$, or a potential term domination leading to a constant density with $w_\phi=-1$.

%%%%%%%%%%%%%%%%%%%%%%%%%%%%%%%%%%%

\subsection{Complex scalar field: Spintessence}

\begin{figure}[t!]
\begin{center}
    \includegraphics[width=12cm]{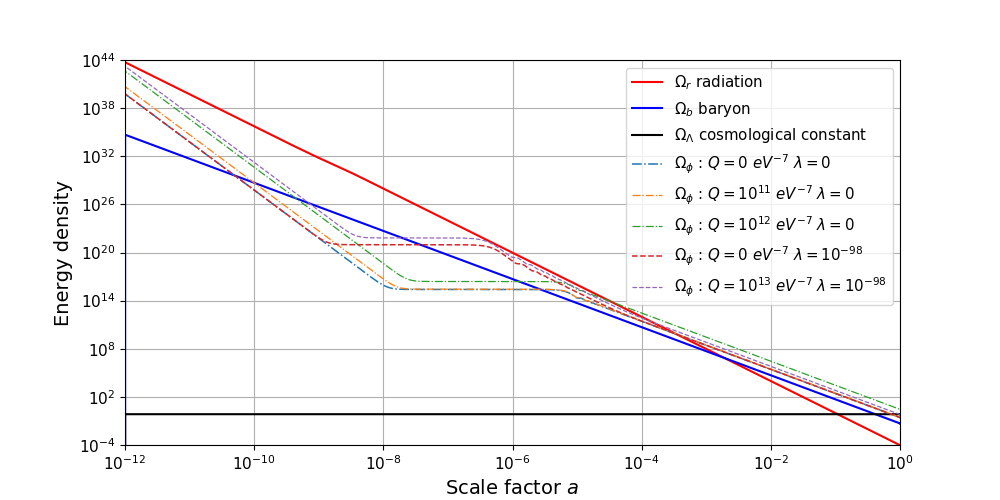}
    \caption{Evolution of a spintessence complex scalar field with a mass term behaving like dark matter, as a function of the scale factor. Baryon, radiation and cosmological constant densities follow the standard values of the $\Lambda$CDM model. We have chosen $\dot{\phi}=1.3 \times 10^{17}$ eV$^{-3}$ for $a=10^{-12}$, and we set the initial value of $\phi$ in order to obtain a matter-like behaviour starting before the observed $z_{eq}$ of Table~\ref{tableau} and a present scalar field density equal to the observed cold dark matter density.\label{fig:4}}
\end{center}
\end{figure}

Until now we have only considered the case of a real scalar field. In this subsection we study a complex scalar field with a $U(1)$-symmetric potential \cite{Boyle:2001du}. The potential will be given by $U(\phi^{\dagger}\phi)$. For a complex scalar field:
\begin{equation}
\phi(t)=\dfrac{\sigma(t)}{\sqrt{2}}\,\exp(i\theta(t))\,,
\end{equation} 
where $\sigma(t)$ is the amplitude and $\theta(t)$ the phase, the system of equations~(\ref{KG_E}) is replaced by: 
\begin{equation}
\begin{aligned}
&\frac{d^2\sigma}{dt^2}+3H\frac{d\sigma}{dt}+U'\left(\frac{\sigma^2}{2}\right)\sigma - \omega^2 \sigma=0 \,,\\
&\frac{d\omega}{dt}\sigma+3H\omega\sigma+2\omega\frac{d\sigma}{dt}=0 \,,\\
&H^2=H^2_0\left(\Omega^0_b a^{-3}+ \Omega_r^0 a^{-4}+\frac{\rho_\phi}{\rho^{c}}\right) \,,
\end{aligned}
\end{equation}
with $\omega(t)=d\theta/dt$ and
\begin{equation}
\rho_\phi = \frac{\dot{\sigma}^2}{2}+\frac{\omega^2\sigma^2}{2} +U\left(\frac{\sigma^2}{2}\right)\,.\label{spint_density}
\end{equation}
The first two terms of the density constitute the kinetic part. The second equation is the imaginary part of the Klein-Gordon equation and implies the conservation of the U(1)--charge per comoving volume $Q=\omega\sigma^2a^3$. We can rewrite the first equation and the scalar field density as:
\begin{equation}
\begin{aligned}
&\frac{d^2\sigma}{dt^2}+3H\frac{d\sigma}{dt}+U'\left(\frac{\sigma^2}{2}\right)\sigma - \frac{Q^2}{\sigma^3a^6}=0\,,\\
&\rho_\phi = \frac{\dot{\sigma}^2}{2}+\frac{Q^2}{2\sigma^2a^6}+U\left(\frac{\sigma^2}{2}\right)\,.\label{KG_E_comp}
\end{aligned}
\end{equation}
We consider now the evolution of the scalar field density for a polynomial potential of order~4:
\begin{equation}
    U(\phi^{\dagger}\phi)=U_0+m^2\phi^{\dagger}\phi+\lambda(\phi^{\dagger}\phi)^2=\frac{1}{2}m^2\sigma^2+\frac{1}{4}\lambda\sigma^4\,.
\end{equation}
Figure~\ref{fig:4} shows the evolution of the scalar field density for different values of the charge per comoving volume $Q$ as a function of the scale factor. We consider first the case where the self-coupling $\lambda$ is zero. When $Q$ is zero we find the exact same result of the case of the real scalar field. In the early Universe, when $Q$ increases the Klein-Gordon equation gives the conservation of $a^3d\sigma/dt$. At this time, the kinetic terms which depend on the values of $\dot{\sigma}^2$ and $Q^2$ dominates, giving an equation of state $w_\phi = 1$. The behaviour then changes and the density becomes constant. We can see that the height of the plateau increases with the charge per comoving volume. The transition between the plateau and the dark matter behaviour, which is given by $H\simeq m$ for $Q=0$, will also increase. In conclusion, the second term in Eq.~(\ref{spint_density}) acts as an extra mass term and the evolution of the complex scalar field density has the same behaviour as a real scalar field. Similar conclusions are obtained for a non-zero self-coupling, and the scalar field has a negligible density at the time of BBN.\\
\\
To summarize, the density of dark matter scalar fields is generally dominated by the kinetic part at the epoch of BBN, leading to an equation of state $w_\phi=1$ and an evolution such that $\rho_\phi \propto a^{-6}$. However, the density is generally negligible at this time. We also saw that non-standard potentials such as polynomials can lead to other behaviours. It is however important to keep in mind that a scalar field with a heavy mass term can behave like dark matter even during the BBN epoch, having an average equation of state $w_\phi=0$ and an evolution such that $\rho_\phi \propto a^{-3}$.

%%%%%%%%%%%%%%%%%%%%%%%%%%%%%%%%%%%
%%%%%%%%%%%%%%%%%%%%%%%%%%%%%%%%%%%

\section{Dark fluid scalar fields}

\begin{figure}[t!]
\begin{center}
\includegraphics[width=12cm]{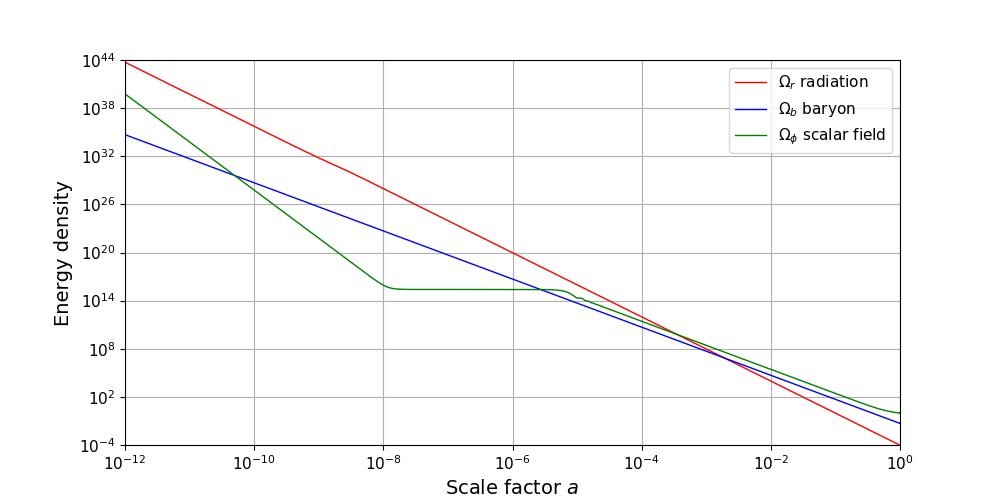}
\caption{Evolution of the dark fluid scalar field with a simple potential, aimed at replacing both dark matter and dark energy. Baryon, radiation and cosmological constant densities follow the standard values of the $\Lambda$CDM model. We have chosen $\phi=1.5\times 10^{29}$ eV$^{-4}$ and $\dot{\phi}=1.3\times 10^{17}$ eV$^{-3}$ for $a=10^{-12}$. The initial value of $\phi$ has been chosen in order to have a dark matter behaviour at the time of recombination.\label{fig:1b}}
\end{center}
\end{figure}

In dark fluid models, dark energy and dark matter are considered to be different manifestations of a unique cosmological component called dark fluid. In Ref.~\cite{Arbey:2005fn} constraints from supernov\ae of type Ia, cosmic microwave background, large scale structures on such models are considered. In this section, we will study the background evolution of a dark fluid scalar field. We choose the parameters and initial conditions so that the present density of the scalar field is equal to the sum of the cold dark matter and cosmological constant densities, and that the scalar field density behaves like the cold dark matter density at the recombination epoch.

We consider here a simple dark fluid potential, which is the sum of fuzzy dark matter potential and a cosmological constant:
\begin{equation}
    U(\phi)=U_0+\frac{1}{2}m^2\phi^2.
\end{equation}
The constant potential $U_0$ can give a dark energy behaviour when $U_0=\rho_{cr}\Omega_{\Lambda}$ and the quadratic potential can provide the dark matter behaviour. The dark fluid density appears as the sum of the fuzzy dark matter and cosmological constant densities. $U_0$ is therefore fixed to the cosmological constant value, and the mass $m$ is in the range of $10^{-24}-10^{-23}$ eV to be in agreement with galactic observations. Figure~\ref{fig:1b} illustrates the evolution of the scalar field energy density. At the time of BBN, the scalar field is dominated by its kinetic term, giving an equation of state $w_\phi=1$. We checked that once the parameters are fixed to comply with the assumptions of the model and the astrophysical and cosmological data, only marginal changes are possible to the scenario described in the figure. The scalar field density will therefore have a negligible influence at the BBN epoch. On the other side, since the scalar field also has a dark energy behaviour today, it means that if the potential contains quintessence potential terms, results similar to the ones of dark energy scalar fields described in Section~\ref{sect:dark_energy} can also be obtained.

%%%%%%%%%%%%%%%%%%%%%%%%%%%%%%%%%%%
%%%%%%%%%%%%%%%%%%%%%%%%%%%%%%%%%%%

\section{Decaying scalar fields}

\begin{figure}[t!]
\hspace{-2.7cm}
\begin{minipage}[c]{0.63\textwidth}
\includegraphics[width=11cm]{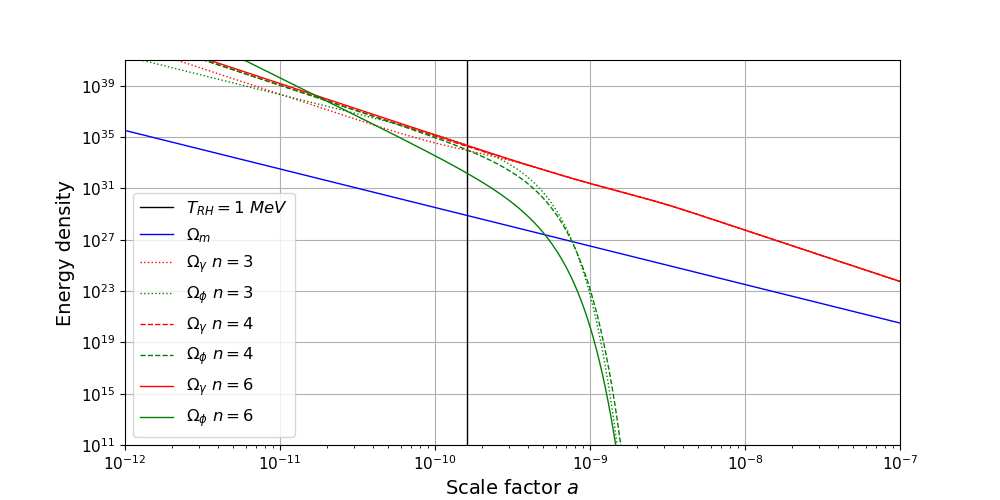}
\end{minipage}
\hspace{0.2cm}
\begin{minipage}[c]{-0.3\textwidth}
\includegraphics[width=11cm]{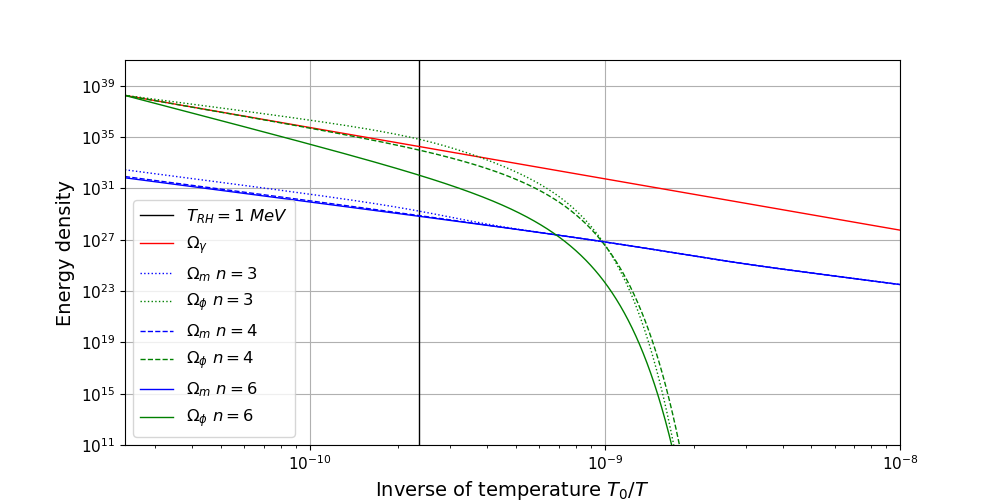}
\end{minipage}
\caption{Evolution of the decaying scalar field density, matter density and radiation density as functions of the scale factor (left) and of the temperature (right), for a reheating temperature of 1 MeV displayed by the vertical black lines. In absence of decay, the scalar field density evolves as $a^{-n}$. The initial densities of the scalar field have been chosen so that $\rho_\phi=\rho_\gamma$ at $T=10$ MeV for different values of $n$.\label{fig:5}}
\end{figure}

In the early Universe, cosmological scalar such as inflatons \cite{Bezrukov:2007ep,Bezrukov:2010jz}, dilatons \cite{deCarlos:1993wie,Gasperini:1994xg}, moduli \cite{Dine:1995uk,Banks:1995dt,Moroi:2001ct,Nakamura:2006uc}, ..., could have existed and given birth to specific phenomena such as inflation or leptogenesis. Their common features is that do not exist any more, at least in non-negligible proportions. We can therefore assume that these scalar fields decayed at a period prior to recombination, so that they did not imprint the CMB. 

Contrary to the cases studied in the previous sections, decaying scalar fields are not constrained by observations of the present Universe, and generally neither by observations of the CMB. Any potentials are therefore possible, with any behaviours. If such scalar fields decay much before BBN, they will have no effect either on BBN or recombination. If they decay after BBN, either their densities are negligible and they have no effect on the cosmological history, or their densities are high enough to impact BBN by modifying the expansion rate and the decay has to occur soon enough in order to escape constraints from CMB. In the latter case, the scalar fields can be considered as stable during BBN, with a constant equation of state $w_\phi \in [-1,1]$, meaning a density scaling as $\rho_\phi \propto a^{-n}$, with $n\in[0,6]$.

We consider now the case of a scalar field decaying at the BBN epoch. To simplify the analysis, we assume an instantaneous thermalization with the thermal bath, and a dominant decay into radiation.

The Klein-Gordon equation becomes:
\begin{equation}
 \Ddot{\phi}+3H\dot{\phi}+\frac{dU}{d\phi}= - \Gamma_\phi \rho_\phi \,,
\end{equation}
and the total radiation entropy receives an injection such that:
\begin{equation}
\frac{{\partial}s_\text{rad}}{{\partial}t} = -3 H s_\text{rad} + \frac{\Gamma_\phi \rho_\phi}{T}\,,
\end{equation}
where $\Gamma_\phi$ is the decay width of the scalar field.

We assume here that the scalar field potential is a power law, such as during BBN and in absence of decay, it evolves as
\begin{equation}
 \rho_\phi = \rho_\phi^0 a^{-n}\,,
\end{equation}
where $n$ is a constant parameter between 0 and 6. The Klein-Gordon equation can thus be rewritten as:
\begin{equation}
 \frac{{d}\rho_\phi}{{d}t} = - n H \rho_\phi  - \Gamma_\phi \rho_\phi\,.
\end{equation}

Following \cite{Gelmini:2006pw,Arbey:2018uho} we define the reheating temperature $T_\text{RH}$ of the scalar field as:
\begin{equation}
 \Gamma_\phi = \sqrt{\frac{4\pi^3 g_\text{eff}(T_\text{RH})}{45}} \, \frac{T^2_\text{RH}}{M_P}\,,
\end{equation}
where $g_\text{eff}$ is the number of effective energy degrees of freedom of radiation. The model is therefore defined by three parameters: the exponent $n$, the reheating temperature $T_\text{RH}$ and the initial scalar field density. 

Figure~\ref{fig:5} shows the evolution of the scalar field as a function of the scale factor or the temperature, for a reheating temperature of 1~MeV. The scalar field density at the initial temperature $T_i=10$ MeV has been chosen so that $\rho_\phi=\rho_\gamma$ for $n=3,4,6$. Since the decay of the scalar field induces a reheating by increasing the radiation density, the initial value of the matter density has been adjusted in order to obtain the observed baryon-to-photon ratio from the CMB. This effect can be seen in the left panel of the figure, where the evolution of the densities is shown as a function of the scale factor: the radiation density increases at a scale factor corresponding to the reheating temperature. This is the reason why the photon density is different for each choice of $n$. In the right panel where the evolution is shown as a function of the temperature, this effect translates into a decrease of the matter density, which is equivalent to lowering the baryon over radiation ratio to obtain the value derived from the CMB observations. This is also the reason why the matter density appears as different for each choice of $n$.

We see that after reheating, the decrease of the scalar field accelerates and its density drops very quickly, increasing simultaneously the radiation density. It can therefore be expected that a decaying scalar field can modify BBN for $T_{\rm RH} \in [1,10]$ MeV.

%%%%%%%%%%%%%%%%%%%%%%%%%%%%%%%%%%%
%%%%%%%%%%%%%%%%%%%%%%%%%%%%%%%%%%%

\section{BBN constraints on cosmological scalar fields}

In the previous sections, we have studied the cosmological behaviours of a broad range of scalar fields. We have seen that stable and decaying scalar fields have different consequences and study them separately.

\subsection{Stable scalar fields}

\begin{figure}[p]
\vspace{-1cm}
\hspace{-2.5cm}
\begin{minipage}[c]{0.63\textwidth}
	\includegraphics[width=10cm]{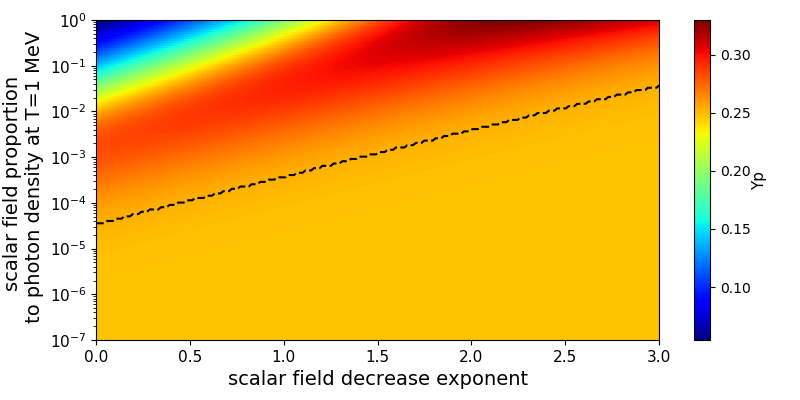} 
    \includegraphics[width=10cm]{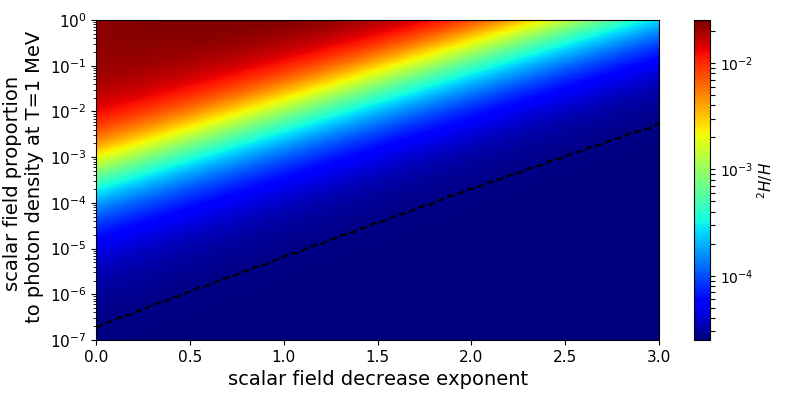} 
    \includegraphics[width=10cm]{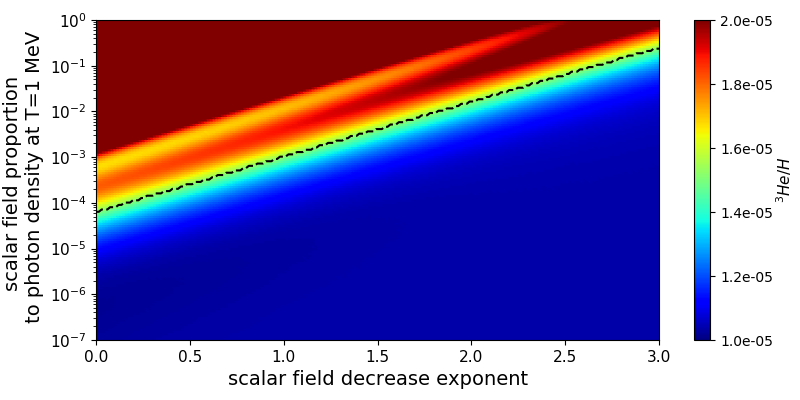} 
    \includegraphics[width=10cm]{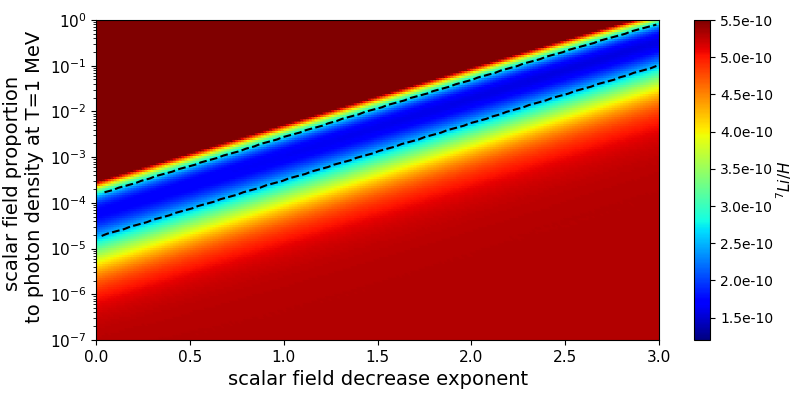} 
\end{minipage}
\hspace{0.2cm}
\begin{minipage}[c]{-0.3\textwidth}
    \includegraphics[width=10cm]{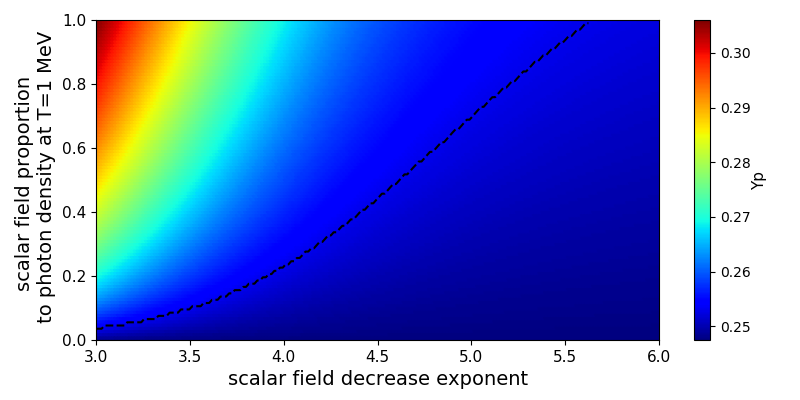}
    \includegraphics[width=10cm]{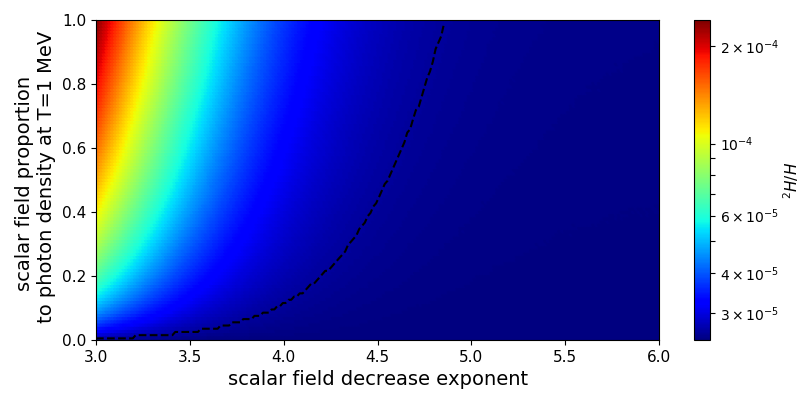} 
    \includegraphics[width=10cm]{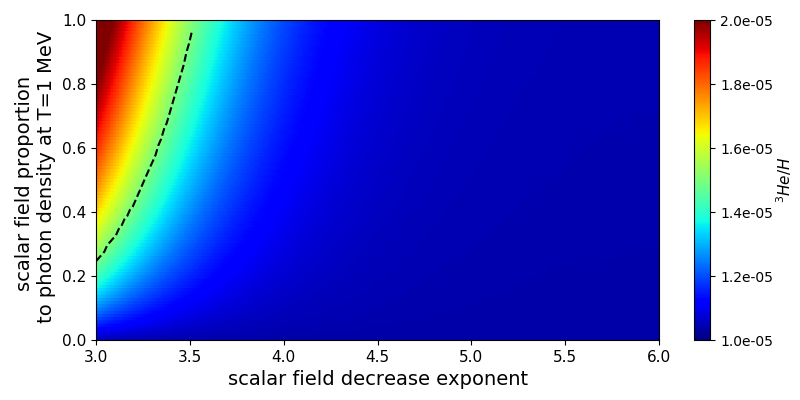} 
    \includegraphics[width=10cm]{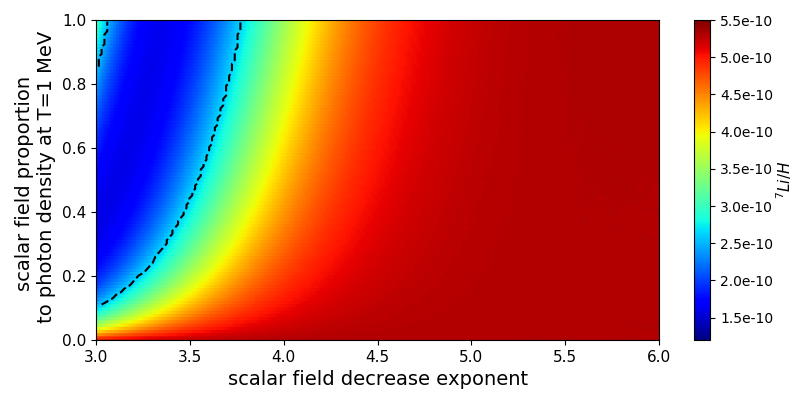}
\end{minipage}
    \caption{From top to bottom, values of $Y_p$, $^2H/H$, $^3He/H$ and $^7Li/H$, as a function of the decrease exponent $n$ and the initial scalar field density (normalized to photon density) at $T=1$ MeV, (left) with a logarithm scale on the y-axis and for $n\le3$, and (right) with a linear scale on the y-axis and for $n\ge 3$. The dashed lines represent the individual BBN constraints at 95\% C.L. Except for $^7Li$, the excluded region is above the lines.\label{fig_alter_standmod}}     
\end{figure}

Stable scalar fields have in general a constant equation of state during BBN, with $w_\phi$ ranging from $-1$ to $+1$ corresponding to a density varying as $a^{-n}$ with $n\in[0,6]$. The most standard values are 1 (kinetic term domination) and 1/3 (radiation-like behaviour), corresponding to densities scaling as the inverse of the scale factor with an exponent 6 and 4, respectively. 

In the cases when $w_\phi<0$, since the scalar field density decreases more slowly than radiation, if the scalar field density is not completely negligible, it will start dominating the expansion of the Universe during or after BBN, and can even affect the CMB.

In presence of a scalar field, the expansion rate given by the Friedmann equation will be modified during BBN. Two extreme cases can be considered. The first one corresponds to a scalar field density negligible with respect to the photon density at $T\sim1$ MeV. We will obtain in this case the same results as in the standard cosmological model. The opposite case is the domination of the scalar field density at BBN time, which will modify the abundance of the elements via Hubble rate modification.

Figure~\ref{fig_alter_standmod} shows $Y_p$ and the abundances of ${}^2H$, ${}^3He$ and ${}^7Li$ for different values of the exponent $n$ and the scalar field density at $T\sim1$ MeV. The dashed lines correspond to exclusions by the individual constraints for each element at $95\%$ C.L. Using the $\chi^2$ approach we obtain at $95\%$ C.L.:
\begin{itemize}
\item for $n_{\phi}=0$ (constant density, potential domination), BBN constraints exclude:
\begin{equation}
\rho_{\phi} (1\;{\rm MeV}) \gtrsim 2\times 10^{-7}\,\rho_{\gamma}(1\;{\rm MeV})\,.
\label{C0}
\end{equation}
\item for $n_{\phi}=3$ (matter-like behaviour), BBN constraints exclude:
\begin{equation}
\rho_{\phi} (1\;{\rm MeV}) \gtrsim 0.005\,\rho_{\gamma}(1\;{\rm MeV})\,.
\label{C1}
\end{equation}
\item for $n_{\phi}=4$ (radiation-like behaviour), BBN constraints exclude:
\begin{equation}
\rho_{\phi}(1\;{\rm MeV}) \gtrsim 0.11\,\rho_{\gamma}(1\;{\rm MeV})\,.
\label{C2}
\end{equation}
\item for $n_{\phi}=6$ (kinetic term domination), BBN constraints exclude:
\begin{equation}
\rho_{\phi}(1\;{\rm MeV}) \gtrsim 1.40\,\rho_{\gamma}(1\;{\rm MeV})\,.
\label{C3}
\end{equation}
\end{itemize}

We also see that the ${}^7Li$ abundance can be modified by the scalar field, but as we can see from Fig.~\ref{fig_alter_standmod}, there is no region where the predictions for $Yp$, ${}^2H$, ${}^3He$ and ${}^7Li$ can be simultaneously in agreement with their observational values.

In terms of physical models, scalar fields behaving like matter (fuzzy dark matter, ...) do not affect BBN and no constraint can be found. For quintessence or dark field models on the other hand, Eq.~(\ref{C2}) applies for tracking scenarios with fixed points solutions, or Eq.~(\ref{C3}) in more general scenarios with generic initial conditions.
For more exotic scenarios, BBN constraints can be obtained from Fig.~\ref{fig_alter_standmod}.

%%%%%%%%%%%%%%%%%%%%%%%%%%%%%%%%%%%

\subsection{Decaying scalar fields}

We now turn to the case of decaying scalar fields. As discussed before, if the scalar field decays after BBN, the constraints that we just obtained as still valid. On the other hand, if the decay is finished when BBN starts, the scalar field will have a negligible density which will have no effect on BBN.

We now turn to the case of decaying scalar fields with reheating temperatures below 10 MeV. We consider scalar fields which evolve as $\rho_\phi \propto a^{-n}$ in absence of decay, for $n=0,3,4,6$.

\begin{figure}[p]
\vspace{-1cm}
\hspace{-2.5cm}
\begin{minipage}[c]{0.63\textwidth}
	\includegraphics[width=10cm]{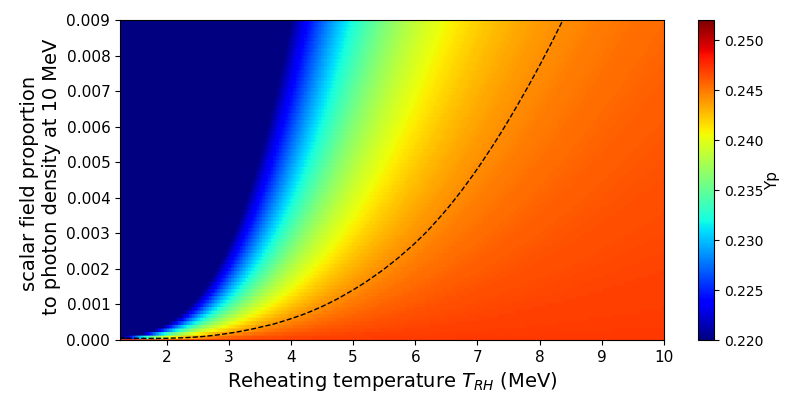}
    \includegraphics[width=10cm]{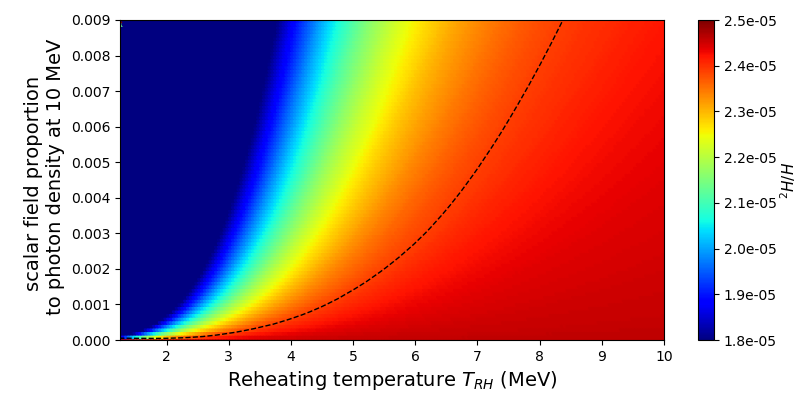}
    \includegraphics[width=10cm]{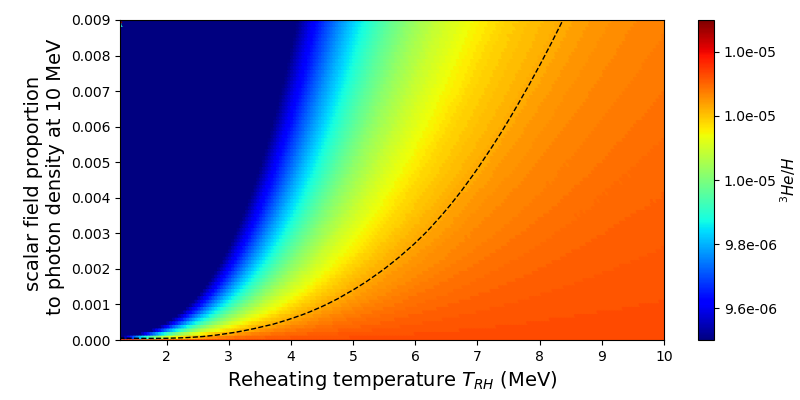}
    \includegraphics[width=10cm]{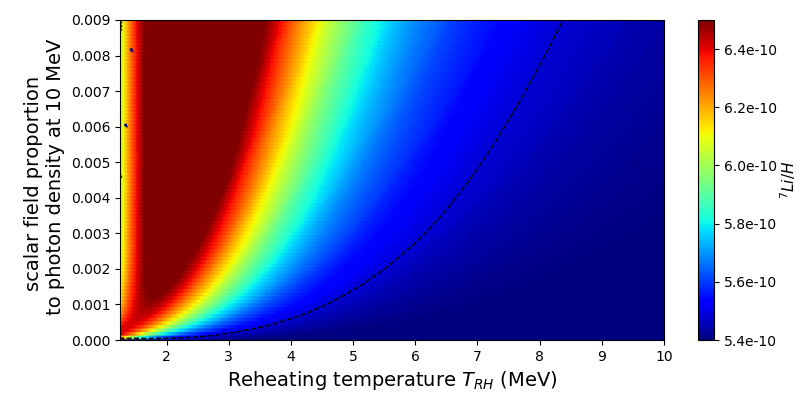}
\end{minipage}
\hspace{0.2cm}
\begin{minipage}[c]{-0.3\textwidth}
    \includegraphics[width=10cm]{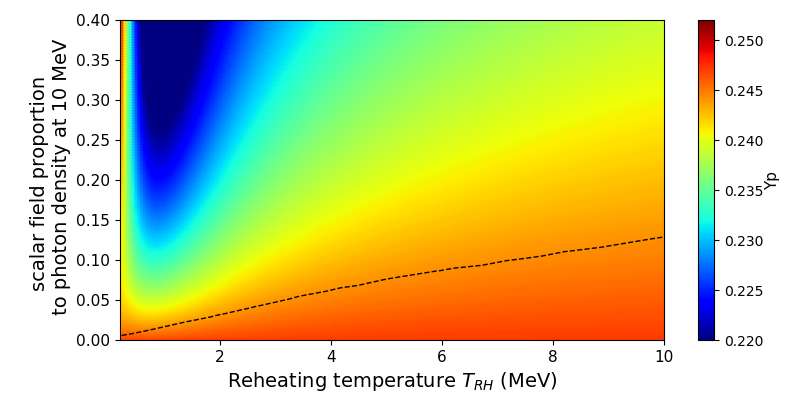}
    \includegraphics[width=10cm]{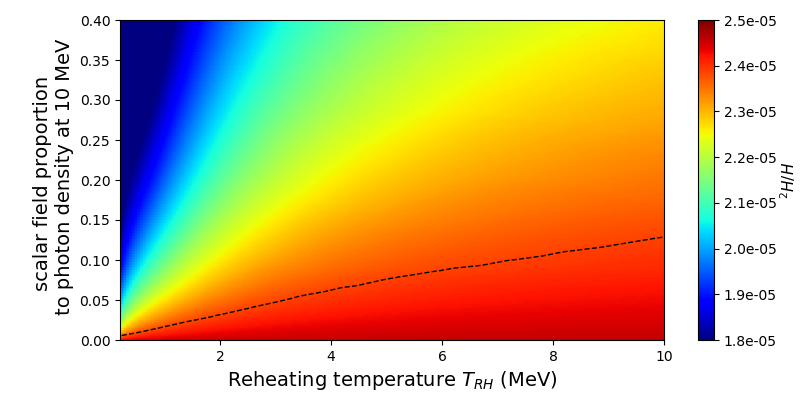}
    \includegraphics[width=10cm]{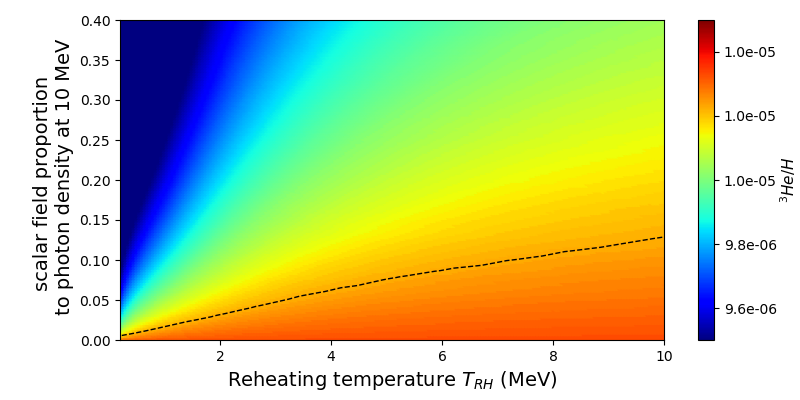}
    \includegraphics[width=10cm]{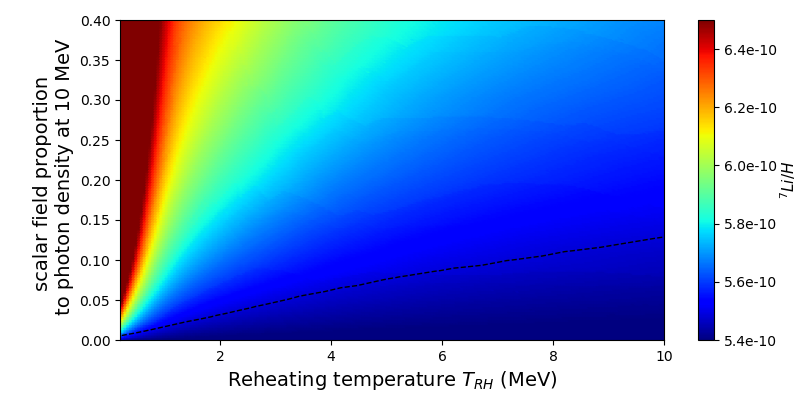}
\end{minipage}
    \caption{From top to bottom, values of $Y_p$, $^2H/H$, $^3He/H$ and $^7Li/H$, as functions of the reheating temperature $T_{\rm RH}$ and the initial scalar field density (normalized to photon density) at $T=10$ MeV, for (left) a constant density $n=0$, and (right) a matter-like behaviour $n=3$. The dashed lines represent the individual BBN constraints at 95\% C.L. Except for $^7Li$, the excluded regions are above the lines.}
    \label{fig_alter_phi03}
\end{figure}

\begin{figure}[p]
\vspace{-1cm}
\hspace{-2.5cm}
\begin{minipage}[c]{0.63\textwidth}
	\includegraphics[width=10cm]{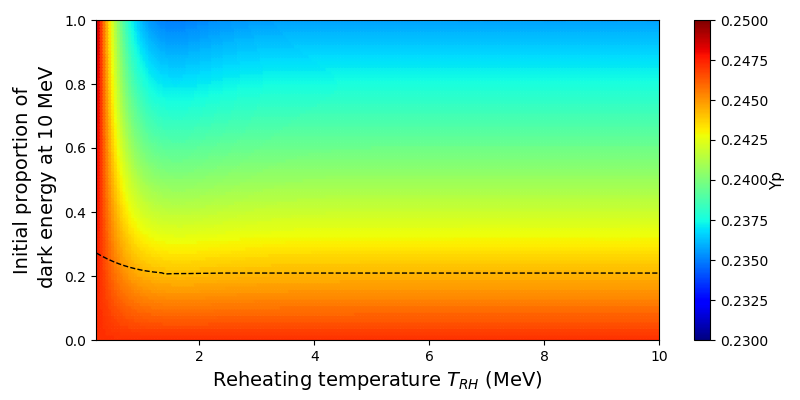}
    \includegraphics[width=10cm]{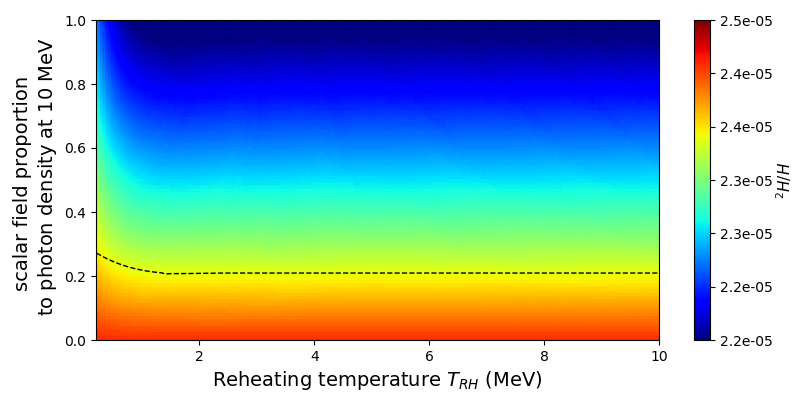}
    \includegraphics[width=10cm]{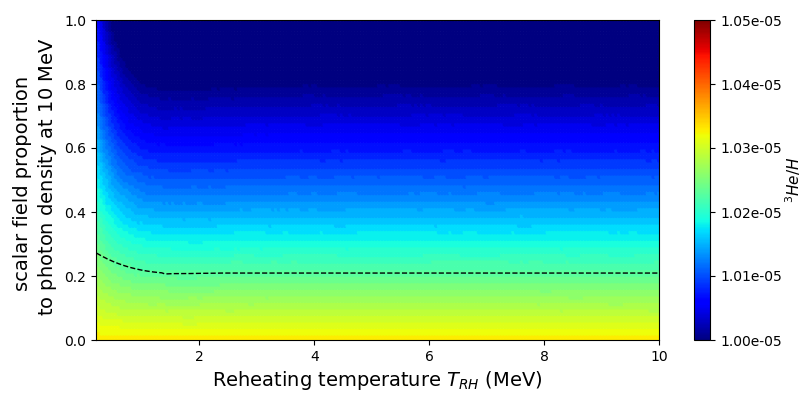}
    \includegraphics[width=10cm]{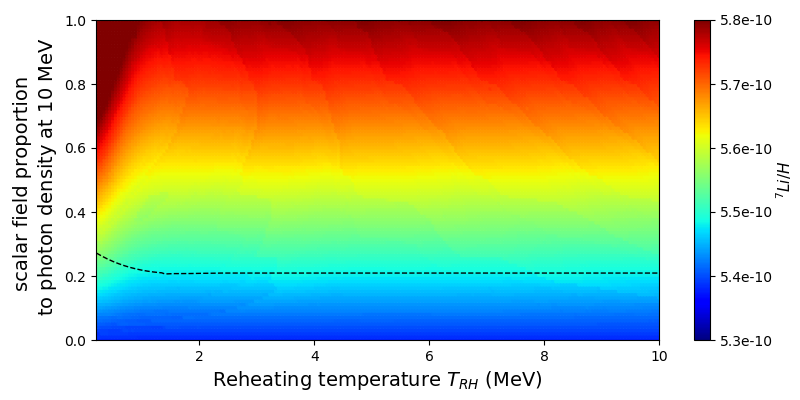}
\end{minipage}
\hspace{0.2cm}
\begin{minipage}[c]{-0.3\textwidth}
    \includegraphics[width=10cm]{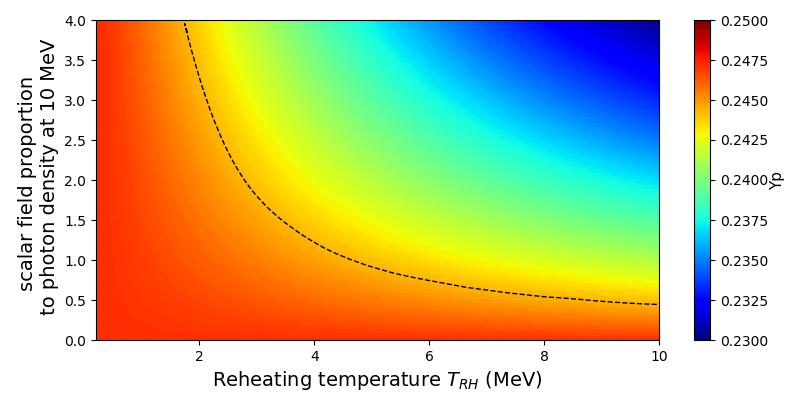}
    \includegraphics[width=10cm]{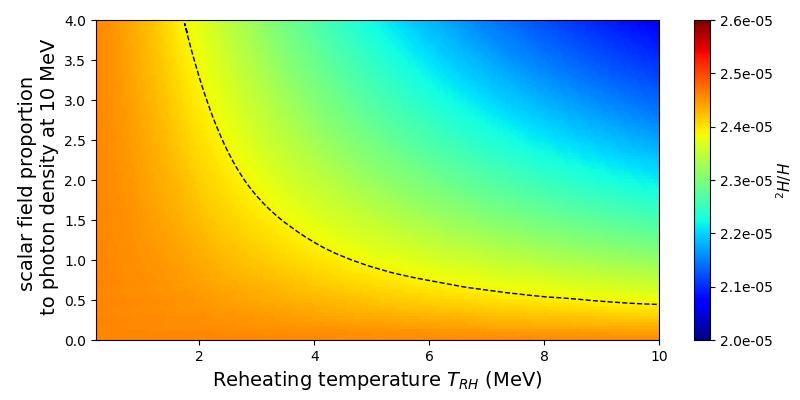}
    \includegraphics[width=10cm]{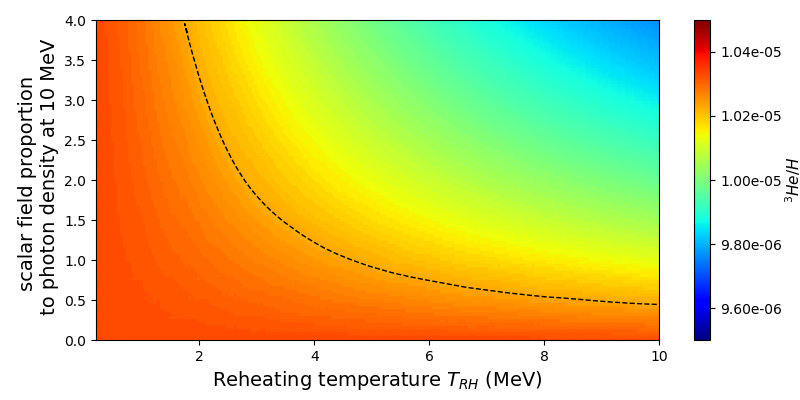}
    \includegraphics[width=10cm]{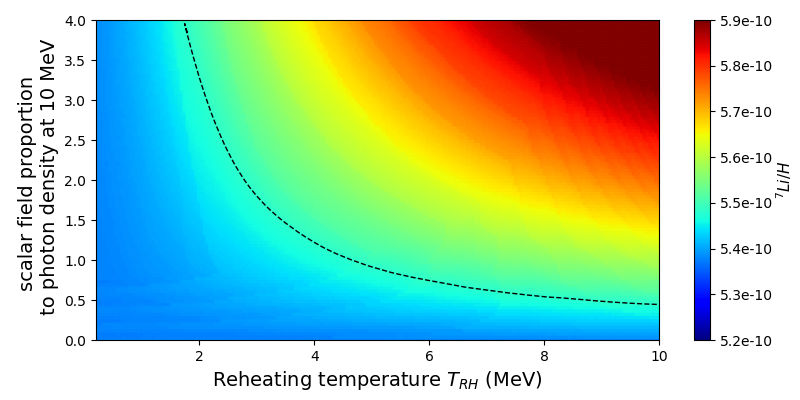}
\end{minipage}
    \caption{From top to bottom, values of $Y_p$, $^2H/H$, $^3He/H$ and $^7Li/H$, as functions of the reheating temperature $T_{\rm RH}$ and the initial scalar field density (normalized to photon density) at $T=10$ MeV, for (left) a radiation-like behaviour $n=4$, and (right) a dominating kinetic term $n=6$. The dashed lines represent the individual BBN constraints at 95\% C.L. Except for $^7Li$, the excluded regions are above the lines.}
    \label{fig_alter_phi46}
\end{figure}

In Figure~\ref{fig_alter_phi03}, we present the abundance of the elements for $n=0,3$ as functions of the reheating temperature $T_{\rm RH}$ and the ratio of the scalar field density to the photon density at $T=10$ MeV\footnote{It is important to note that the initial densities of the decaying scalar fields are fixed at $T=10$ MeV, contrary to the case of stable scalar fields where the initial temperature was $T=1$ MeV.}. The same plots are obtained with $n=4,6$ in Fig.~\ref{fig_alter_phi46}. The dashed lines correspond to exclusions by the individual constraints for each element at $95\%$ C.L. Using the $\chi^2$ approach at 95\% C.L. we obtain the following results:
\begin{itemize}
\item for $n=0$, the compatible parameter region corresponds to large reheating temperatures and small densities. Indeed, before the decay, the density of the scalar field is constant, meaning that it will dominate and accelerate the expansion over the radiation density, which decreases as $T^4$. This model is therefore strongly constrained, so that only a small fraction of constant scalar field can be allowed.

\item for $n=3$, since the scalar field has a matter-like behaviour, if the scalar field density dominates the expansion, the expansion rate is smaller before its decay. We see that such a scenario is excluded if
\begin{equation}
\rho_{\phi}(10\;{\rm MeV}) \gtrsim 0.01 \,\left(\frac{T_{\rm RH}}{1\;{\rm MeV}}\right) \, \rho_{\gamma}(10\;{\rm MeV})\,.
\end{equation}
There is therefore a compensation between the modification of the expansion rate and the injection of radiation.

\item for $n=4$, we obtain a limit
\begin{equation}
\rho_{\phi}(10\;{\rm MeV}) \gtrsim 0.1\,\rho_{\gamma}(10\;{\rm MeV})\,,
\end{equation}
which is equivalent to the limit obtained for a stable scalar field. This could be expected since the scalar field had originally a radiation-like behaviour and decays into radiation.

\item for $n=6$, BBN is unaffected for $\rho_{\phi}(10\;{\rm MeV}) \lesssim 0.5\,\rho_{\gamma}(10\;{\rm MeV})$ independently of the reheating temperature, or when the reheating temperature is below 4 MeV and $\rho_{\phi}(10\;{\rm MeV}) \lesssim \rho_{\gamma}(10\;{\rm MeV})$. For large reheating temperatures, the scalar field decays earlier, increases the radiation density, decelerates the expansion, and modifies the abundance of the elements even for small values of the initial density. For lower reheating temperatures, the expansion rate is increased but slowed before the decay, and less constraining limits are thus obtained.

\end{itemize}

Similarly to the case of stable scalar fields, there is no possibility to simultaneously explain the abundance of $^7Li$ and be consistent with the constraints on the $^4He$, $^2H$ and $^3He$ abundances.

To summarize, in the very early Universe the primordial scalar fields are likely to have a constant density or a dominating kinetic term. For the constant behaviour, it is mandatory for the scalar field to have a subleading density at BBN time, independently of the reheating temperature. On the contrary, for a kinetic term dominated scalar field, the initial density can be rather large, and even dominant for low reheating temperature, meaning that a late reheating is favoured in such cases.

%%%%%%%%%%%%%%%%%%%%%%%%%%%%%%%%%%%
%%%%%%%%%%%%%%%%%%%%%%%%%%%%%%%%%%%

\section{Conclusions}

In this paper, we have studied the cosmological evolution of scalar fields and studied their impact on Big-Bang nucleosynthesis. We have shown that these can have non-negligible densities $\rho_\phi$ at the time of BBN and equations of state $w_\phi=P_\phi/\rho_\phi$ between $-1$ and $+1$.

Scalar fields can replace a cosmological constant, for example in quintessence models, and we saw that the most usual dark energy scenarios led to $w_\phi=0$, $w_\phi=1/3$ or $w_\phi=1$ at BBN time, which can affect BBN if the scalar field density is non-negligible at this epoch.

Scalar fields can also act as dark matter, with $w_\phi=0$ today, as it is the case in fuzzy dark matter scenarios. In the early Universe, the kinetic term generally dominates, giving an equation of state $w_\phi=1$ at BBN time. However, because of the constraints from studies of the CMB, the density of such scalar fields is negligible at the time of BBN, and no constraint can be obtained from BBN.

Similarly, dark fluid models with scalar fields replacing simultaneously dark matter and dark energy are extremely constrained both by dark matter constraints at local and large scales and by dark energy constraints at cosmological scales, and we showed that in the most simple models, no constraints can be obtained from BBN since the scalar field density at BBN epoch is negligible. Only more complex models incorporating specific dark energy potentials are likely to have effects on BBN, and the constraints  are expected to be similar to the ones obtained on quintessence scenarios.

Primordial scalar fields which have decayed during BBN are on the contrary more likely to have affected the abundance of the elements, in two different ways. First the scalar field density increases the total density and affects the expansion rate of the Universe. This effect can be particularly important since no strong constraint can limit the decaying scalar field parameters at BBN time, so that large densities are still possible during BBN. Second the decay into radiation injects entropy which modifies the relation between time (or scale factor) and temperature and generates a reheating at the BBN epoch. We considered the most usual cases, {\it i.e.} a scalar field density scaling as $a^{-n}$ in absence of decay, with $n=0,3,4,6$, and derived constraints on the reheating temperature $T_{\rm RH}$ and the initial scalar field density. 

Whichever scalar field model is used, we showed that it is not possible to find set-ups which simultaneously satisfy the $Y_p, ^2H, ^3He$ constraints and the $^7Li$ one. In other terms, the lithium problem cannot be solved via the scalar field models that we considered. It may however be possible to design scenarios with scalar fields decaying into specific particles which may affect BBN and decrease the abundance of lithium-7. We defer this task to later studies.

\bibliographystyle{h-physrev5}
\bibliography{biblio}

\end{document}